\documentclass[10pt,preprint]{aastex}
\usepackage{graphics,graphicx}
\shortauthors{GWINN ET AL.}
\shorttitle{Intermittent Emission}
\begin{document}

\title{Effects of Intermittent Emission:\\ Noise Inventory for Scintillating Pulsar B0834$+$06}

\author{C. R. Gwinn, M. D. Johnson}
\affil{Department of Physics, University of California, Santa Barbara, California 93106, USA}
\email{cgwinn@physics.ucsb.edu, michaeltdh@physics.ucsb.edu} 

\author{T.V. Smirnova}
\affil{Pushchino Radio Astronomy Observatory of Lebedev Physical Institute, 142290 Pushchino, Russia}
\email{tania@prao.ru}

\author{D.R. Stinebring}
\affil{Department of Physics and Astronomy, Oberlin College, Oberlin, OH 44074, USA}
\email{dan.stinebring@oberlin.edu} 

\vskip 1 truein
\begin{abstract}
We compare signal and noise for observations of the scintillating pulsar B0834+06, using very-long baseline interferometry and a single-dish spectrometer. Comparisons between instruments and with models suggest that amplitude variations of the pulsar strongly affect the amount and distribution of self-noise. We show that noise follows a quadratic polynomial with flux density, in spectral observations. Constant coefficients, indicative of background noise, agree well with expectation; whereas second-order coefficients, indicative of self-noise, are $\approx 3$ times values expected for a pulsar with constant on-pulse flux density. We show that variations in flux density during the 10-sec integration accounts for the discrepancy. In the secondary spectrum, $\approx 97\%$ of spectral power lies within the pulsar's typical scintillation bandwidth and timescale; an extended scintillation arc contains $\approx 3\%$. For a pulsar with constant on-pulse flux density, noise in the dynamic spectrum will appear as a uniformly-distributed background in the secondary spectrum. We find that this uniform noise
background contains 95\% of noise in the dynamic spectrum for interferometric observations;
but only 35\% of noise in the dynamic spectrum for single-dish observations. Receiver and sky dominate noise for our interferometric observations, whereas self-noise dominates for single-dish. We suggest that intermittent emission by the pulsar, on timescales $< 300\ \mu{\rm sec}$, concentrates self-noise near the origin in the secondary spectrum, by correlating noise over the dynamic spectrum. We suggest that intermittency sets fundamental limits on pulsar astrometry or timing. Accounting of noise may provide means for detection of intermittent sources, when effects of propagation are unknown or impractical to invert.
\end{abstract}


\clearpage

\section{INTRODUCTION}

\subsection{Noise}\label{sec:dicke}

Astronomical measurements are comprised of a deterministic part, the signal;
and a random part, noise.
The present paper is primarily concerned with noise in observations of the pulsar
PSR B0834$+$06.
Interstellar scintillation is responsible for most, if not all, of the variations of the flux density with frequency;
scintillation and intrinsic variations are responsible for variations with time.
We observed the pulsar using two instruments: with the Wideband Arecibo Pulsar Processor (WAPP), a specialized
single-dish spectrometer, with pulsar gate,
at Arecibo Observatory\footnote{The Arecibo Observatory is part of the National Astronomy and Ionosphere Center, which is operated by Cornell University under a cooperative agreement with the National Science Foundation.}; and with very-long baseline interferometry observations on the Arecibo-Jodrell baseline,
processed with the correlator of the Very Long Baseline Array (VLBA)\footnote{The Very Long Baseline Array is a facility of the National Radio Astronomy Observatory. The National Radio Astronomy Observatory is a facility of the National Science Foundation operated under cooperative agreement by Associated Universities, Inc.}, also with a pulsar gate.
We also observed the pulsar with the BSA telescope at Puschino to determine scintillation parameters.
We match a theoretical model to noise, as measured in various ways, and find the magnitude 
of noise and self-noise for this object.  We compare noise in the dynamic scintillation spectrum
(intensity or interferometric visibility, measured with observing frequency and time),
and in the secondary spectrum (measured in the Fourier-conjugate domain of lag and rate).

Radio-astronomical signals are usually assumed to be intrinsically noiselike:
the observed electric field is drawn from a Gaussian distribution with zero mean,
as are contributions from backgrounds and instruments.
The variances and covariances of the Gaussian 
distribution are the desired, deterministic, signal.
For finite samples, the variance cannot be measured exactly, so that the 
source itself contributes to noise: a
phenomenon know as ``source noise'' or ``self-noise"
\citep{kul89,ana91,viv91,Gwi06}.
The Dicke Equation  
incorporates this effect:
the root-mean-squared error $\delta I$ in a measurement of flux density $I$ is given by \citep{Dic46,Bur01}:
\begin{equation}
(\delta I)^2 = {{I^2}\over{ N_o }} ,
\label{eq:dicke_eq}
\end{equation}
where $N_o = \Delta\nu\;\Delta t$ is the number of independent samples,
for an observed bandwidth $\Delta\nu$ and integration time $\Delta t$.
The analogous expression holds for interferometric visibility \citep{tms01}.
Flux density $I$ is often expressed in units of temperature. In this paper we use Jy, or
instrumental units for the same quantity: VLBA correlator units (V.c.u.) or WAPP units (W.u.).
In Eq.\ \ref{eq:dicke_eq}, the flux density $I =I_S + I_n$ includes that of the source $I_S$ 
and the contribution from 
instrument and backgrounds $I_n$.
For our observations of a scintillating pulsar, we expect spectrally- and temporally-varying signal $I_S$,
and a constant noise background $I_n$.
Varying $I_S$ results in a varying self-noise contribution to $(\delta I)^2$.
For single-dish observations $(\delta I)^2$ is then a quadratic function of $I_S$;
for interferometric observations with 100\% visibility, 
it is a quadratic function of correlated flux density in phase with the signal,
and a linear function, with the same constant and linear coefficients, in quadrature with the signal (\citealp{Gwi06}, see also \S\ref{sec:noise_polynomial} below).

\subsection{Outline}

In this paper we compare 
noise estimated from our observations, by various measures,
with theoretical expectations.
Our observational measures include measuring differences between samples with identical (or nearly identical) signal;
comparison with the distribution of measurements, which includes effects of noise;
and measuring noise in a region empty of signal, under conditions where noise is expected to be stationary.
In
\S\ref{sec:theory} we present theoretical background, notation, and discussion of units.
We explain the origin of the quadratic polynomial that describes noise.
In \S\ref{sec:observations} we describe the observations, and give scintillation parameters for our program pulsar.
In \S\ref{sec:dynspec} we outline the calibration of the data,
from comparison of amplitude variations using the two instruments,
and an estimate of background noise from the following section.

Differencing consecutive measurements in time
provides a measure
of noise, if the signal does not change between samples,
and if noise is uncorrelated between samples.
This approach involves only observed quantities and is relatively independent of the instrument.
It offers the advantage that one can estimate the average signal as well as departures,
and thus estimate noise as a function of signal.
This technique requires a sufficiently large statistical sample,
as well as sufficiently-slowly varying signal.
We form many spectra over short periods within our observation;
the resulting 2D plot flux density with frequency and time is known as the ``dynamic spectrum'' \citep{Bra98}.
We form distributions of differences between consecutive spectra in \S\ref{sec:diffnoise}, and compare results with theoretical expectation,
via the Dicke Equation and related expressions.
Calibration allows comparison of the background noise level $I_n$ with expectations for the instrument,
and of the mean flux density of the source with that measured by others.
Self-noise is independent of calibration, but depends on the number of samples.
We find that self-noise exceeds expectation from Eq.\ \ref{eq:dicke_eq}, for a source with rectangular pulses of constant intensity, by about a factor of 3;
variations in the flux density of the pulsar over the 10-sec span of accumulation of a spectrum can account for the increase in noise.

The global distribution of observed intensity or visibility also shows effects of noise.
The observed intensity (for single-dish observations) or visibility (for interferometry)
is the sum of signal and noise.
For a scintillating pointlike source, in strong scattering, observed 
at one antenna or on a long baseline, the distribution of average signal is exponential \citep{sch68,Gwi98}.
Noise broadens the distribution; because self-noise is greatest at large flux density,
it broadens the distribution more there.
In \S\ref{sec:global_dist} we present fits of the noise model, and an exponential, to the observed distributions.

A region vacant of signal provides a measure of the background noise level.
The ``lag-rate'' correlation function \citep{tms01},
the 2D Fourier transform of the observed dynamic spectrum,
provides many such regions.
The lag-rate function gives the Fourier transform of flux density, with lag (conjugate to frequency) and rate (conjugate to time).
Parseval's Theorem shows that the total noise in the lag-rate correlation function is equal to the mean square noise in the dynamic spectrum.

We find that the lag-rate correlation functions for our observations show a nearly-stationary background noise level.
This noise background contains
95\% of noise for the interferometric data (dominated by instrumental noise),
and 40\% of that expected for the single-dish data (dominated by self-noise).
Thus, we find that self-noise is not uniformly distributed in the lag-rate spectrum:
self-noise must be concentrated in regions where signal is strong.

Background noise from sky and instrument is usually nearly ``white'': it is stationary, and 
individual samples are uncorrelated \citep{Pap91}.
Self-noise is not stationary in the dynamic spectrum,
because the flux density of the source varies with time and frequency.
(Traditionally, the term ``stationary'' indicates that statistical properties are invariant in time;
here, we broaden the term to include invariance in other domains, such as frequency, as well).
However, if noise in adjacent samples of the dynamic spectrum is uncorrelated, then the noise can be described as white noise, times an
envelope that varies with frequency and time.
This envelope is the variance of the noise. 

The Fourier transform of white noise is white noise \citep{Pap91},
and the convolution of white noise with any function is stationary.
Consequently,
if self-noise is uncorrelated between samples in the dynamic spectrum,
then it is expected to be distributed uniformly in the lag-rate correlation function. 
Because we observe that self-noise is not uniformly distributed, we conclude that self-noise in adjacent samples of the dynamic spectrum is correlated.
Such correlation can arise from digitization \citep[][\S 4.3.1]{Gwi06}
and, much more strongly, from intermittent emission at the pulsar \citep{Gwi10}.
Intermittent emission will introduce correlation of self-noise between spectral channels.
We suggest that variations of pulsar flux density over the timescale of integration of one spectrum,
or $\approx 300\ \mu{\rm sec}$, redistribute self-noise in the secondary spectrum.

We summarize our results in \S\ref{sec:summary}.

\subsection{Scintillation and Arcs}

Pulsars and other compact radio sources scintillate in the interstellar plasma, 
because of multipath propagation 
and consequent interference among paths \citep{Ric77}.
This scintillation is random, but exhibits characteristic scales of variation in time and frequency: the scintillation time $t_d$
and scintillation bandwith $\Delta\nu_d$, respectively.
These scales result from a characteristic scattering angle $\theta$ \citep{Shi70,Gwi98}.
These are observed in, and measured from, dynamic spectra: sequential time series of spectra.

Scintillation arcs are a separate, but related phenomena \citep{Hil03}.
They characterize structure in the dynamic spectra on frequency and time scales smaller than $t_d$ and $\Delta\nu_d$.
They thus correspond to angular deflections larger than $\theta$.
Such substructure, on scales smaller than $t_d \times \Delta\nu_d$ has long been observed
\citep{Cor86,Wol87,Gup94,Ric97}.
\citet{Sti01} found that
these substructures are characterized by
narrow parabolic arcs in the ``secondary spectrum,''
the square modulus of the two-dimensional Fourier transform of the dynamic spectrum with time and frequency.
A series of papers has explored the properties and interpretation of these arcs.
Many of these involve observations of PSR B0834$+$06, and interpretation of those
observations.
Among these are \citet{Hil03, Wal04, Cor06, Wal08}.
Although we observe arcs in the secondary spectrum, we focus on noise in this paper.

\section{THEORETICAL BACKGROUND}\label{sec:theory}

\subsection{Noise and Self-Noise}\label{sec:noise_polynomial}

\subsubsection{Electric Field, Visibility, and Intensity}

Noise is the departure of a given measurement of some observable $V$ from the ensemble-average:
$\delta V = V - \langle V \rangle_n$.
Here, the angular brackets $\langle ... \rangle_n$ 
denote an average over an ensemble of statistically-identical measurements with different noise,
but with scintillation spectrum held fixed.
The subscript ``$n$'' indicates that the average is over noise;
we do not average over realizations of the scintillation pattern that is commonly used to study scintillation \citep[see, for example][]{Ric77,Gwi98}, or an average over variations of flux density of the pulsar \citep{Ric75}.
The variance of noise characterizes its magnitude:
\begin{equation}
\langle \delta V^2 \rangle_n = 
\langle V^2 \rangle_n - \langle V \rangle_n^2 .
\end{equation}
In this section, we motivate the fact that noise varies with signal as a quadratic polynomial,
for both interferometric visibility and intensity, under the simple circumstances of our observations.
We present more detailed mathematical treatments elsewhere \citep{Gwi06,Gwi10}.

The visibility is the averaged product of electric fields at two antennas.
The electric field at any instant is a real quantity;
this is conveniently converted to a complex time series by the electronic equivalent of adding the time series and $i=\sqrt{-1}$ times its Hilbert transform \citep{Bra98}.
A passband of this signal is shifted to baseband and sampled at the Nyquist rate,
as is usual for radio astronomy \citep[][\S 5.3.2]{Lor04}. 
We consider statistics for a single polarization.
Consider measurements of the resulting complex electric field $E_A$ and $E_B$ at stations $A$ and $B$.
Then, the measured visibility is
\begin{equation}
 V = {{1}\over{N_o}} \sum_{j=1}^{N_o} E_{Aj} E_{Bj}^* ,
\end{equation}
where the time series extends over $N_o$ samples, indicated by the index $j$.
These electric fields are superpositions of signal $s$, which we assume to be identical at the two antennas here,
and distinct background noise $n_{A}$ and $n_{B}$.
All of these fields are assumed to be random variables drawn from Gaussian distributions with zero mean.
The variances of these fields are the flux densities of signal and noise:
\begin{eqnarray}
\langle s s^* \rangle_n &=& I_S \\
\langle n_{A} n_{A}^* \rangle_n &=& I_{nA} \nonumber \\
\langle n_{B} n_{B}^* \rangle_n &=& I_{nB} \nonumber 
\end{eqnarray}
Signal and noise electric fields are all uncorrelated.
We assume here that both signal and noise are stationary with time and frequency;
as we discuss in \S\ref{sec:intermittent_emission_self_noise} below, 
this assumption appears not to hold for signal.
The model presented here represents a simple case for comparison, 
and has the same characteristics as the more complicated model required for a non-stationary field.

\subsubsection{Noise Polynomials}

For a short baseline observing in the speckle limit,
the average visibility is simply the flux density:
\begin{equation}
\langle V\rangle_n = I_S .
\end{equation}
Here, the ``speckle limit'' indicates observations 
of a source of dimension $L$ much smaller than the angular resolution of the scattering disk, $L<<\lambda/\theta$,
with spectral resolution and integration time both much less than the scales of scintillation:
$\Delta\nu<<\Delta\nu_d$, and $\Delta t<<t_d$ \citep{ng89,Des92}.
A ``short baseline'' yields angular resolution on the sky much less than the angular broadening from scattering;
in this case the correlated flux density is equal to the flux density of the source \citep{Gwi93}.
The noise for interferometer measurements is given by the two expressions:
\begin{eqnarray}
\langle V V^* \rangle_n - \left|\langle V\rangle_n \right|^2 &=& {{1}\over{N_o^2}}\sum_{j,k=1}^{N_o}  \langle (E_{Aj} E_{Bj}^*)(E_{Ak}^* E_{Bk}) \rangle_n 
- \left|\langle V\rangle_n \right|^2
=  {{1}\over{N_o}} \left\{ I_S^2 + (n_{A} + n_{B}) I_S  + n_{A} n_{B}) \right\} \\
\langle V V \rangle_n  - \left(\langle V\rangle_n\right)^2 &=&{{1}\over{N_o^2}}\sum_{j,k=1}^{N_o}  \langle (E_{Aj} E_{Bj}^*)(E_{Ak} E_{Bk}^*) \rangle_n  
- \left(\langle V\rangle_n\right)^2 =  {{1}\over{N_o}} \left\{  I_S^2 \right\}.
\end{eqnarray}
As \citet{Gwi06} showed in more detailed calculations, digitization leaves the form of these equations unchanged,
although it changes the parameters and adds ``quantization noise.''

To express these equations in a more intuitive way,
and to bring them into closer agreement with the analysis below,
we assume that the interferometer phase has been rotated so that the visibility of the source is purely real
at that time and frequency:
$\arg(\langle V\rangle_n)=0$.
If we parametrize the background noise, then the equations take the forms:
\begin{eqnarray}
\langle \delta {\rm Re}[V]^2 \rangle_n &=& b_0 + b_1 V + b_2 V^2   
\label{eq:visibility_noise_polynomial} \\
\langle \delta {\rm Im}[V]^2 \rangle_n &=&  b_0 + b_1 V \nonumber   .
\end{eqnarray}
The constants $b_0$, $b_1$ are the same for both real and imaginary parts.
Note that $b_2 = 1/N_o$, whereas $b_1$ and $b_0$ depend on background noise as well as $N_o$.

The situation is analogous for intensity, but background noise leads to an offset of the distribution of flux density.
Moreover, the intensity is purely real. We find:
\begin{eqnarray}
\langle I\rangle_n &=& I_S + n_A \\
\langle I^2 \rangle_n - (\langle I\rangle_n)^2 &=& {{1}\over{N_o^2}}\sum_{j,k=1}^{N_o}  \langle (E_{Aj} E_{Aj}^*)(E_{Ak}^* E_{Ak}) \rangle_n 
- (\langle I\rangle_n)^2
=  {{1}\over{N_o}} \left\{ I_S^2 + (2 I_n) I_S  + I_n^2 \right\}
\end{eqnarray}
Parametrized in a form similar to visibility, the noise takes the form:
\begin{eqnarray}
\delta I^2 &=& b_0 + b_1 (I-I_n) + b_2 (I-I_n)^2.
\label{eq:intensity_noise_polynomial}
\end{eqnarray}
Here, of course, the parameter $I_n$ represents an offset of the spectrum.
Again, $b_2 = 1/N_o$.
For single-dish observations with constant signal strength and noise, the polynomial is a perfect square:
$b_1 = 2\sqrt{b_2 b_0}$.
This does not hold for intermittent emission, as discussed in \S\ref{sec:intermittent_emission_self_noise} below.
In principle, $I_n^2 = b_0$, although processing of intensity spectra often removes or alters a spectral baseline,
and thus shifts the polynomial;
we treat these two parameters separately below.
Again, digitization leaves the form of this expression unchanged,
while changing the parameters and adding noise.

The distribution of noise need not be Gaussian, and indeed is usually not \citep{Joh10}.
However, when a number of sample spectra are averaged, the distribution approaches a Gaussian, because of the Central Limit Theorem.
Then,
at each element of the dynamic spectrum, in the frequency-time domain, the noise is Gaussian noise times a spectral and time envelope.
This envelope is simply the standard deviation of noise in the frequency-time domain of the dynamic spectrum,
as given by Eq. \ref{eq:visibility_noise_polynomial} or \ref{eq:intensity_noise_polynomial}.
The intensity in those expressions is the ensemble-averaged intensity, over realizations of noise $\langle I\rangle_n$.

\subsection{Amplitude Variations and Self-Noise}\label{sec:intermittent_emission_self_noise}

Pulsars display a rich variety of amplitude variations on many time scales.
Such amplitude variations can be described by a scalar parameter $A(t)$ that multiplies the electric fields $E$.
These variations all affect self-noise.
We divide the variations into 3 regimes.
Longer-term variations are those among different samples of the spectrum; for our observations, these are longer than the 10-sec period of our integrations.
Intermediate-term variations take place on periods shorter than the 10-sec integration time, but longer than the $\approx 300\ \mu$sec period for accumulation of a single spectrum.
Short-term variations take place on periods shorter than accumulation of one spectrum.

\subsubsection{Longer-Term Amplitude Variations}\label{sec:long_term_amp_vars_theory}

Pulse-to-pulse variations in amplitude are described by a gain-like parameter $A$.
If regarded as noise, as for example when differencing consecutive spectra,
these variations have effects similar to the coefficient $b_2$ for self-noise in Eqs. \ref{eq:visibility_noise_polynomial} and \ref{eq:intensity_noise_polynomial}: both introduce a variation that is proportional to the average signal: $\delta I \propto I$.
For interferometric visibility, that variation is in phase with the signal.
The primary distinction between  such amplitude variations and self-noise is that amplitude variations scale the entire spectrum before
the convolution-like addition of noise, rather than at the same time.
We characterize these amplitude variations in \S\ref{sec:amplitude_variations} below,
and consider effects on noise estimates in the subsequent  sections.

\subsubsection{Intermediate-Term Amplitude Variations}\label{sec:intermediate_term_amplitude_variations}

Intermediate-term amplitude variations, among samples averaged together in time, have a direct effect on self-noise,
so that $b_2\geq 1/N_o$.
As an example of such variation, consider variations in amplitude among a series of spectra, before averaging.
One sample of intensity $I_k (t_j)$ is the square modulus of the electric field, in spectral channel $k$ at time $t_j$.
We assume that the electric field is drawn from a Gaussian distribution, and that the draws are uncorrelated for 
different samples of $I_k(t_j)$ \citep{Ric75,Gwi10}.
We suppose that the variance of the electric field changes with time, as described by a gain factor $A(t_j)$.
(For simplicity, we will assume that this gain is the same for all channels; although this is not essential to the argument).
A statistical average over noise recovers the gain: $\langle I_k(t_j)\rangle_n = A(t_j) I_{0k}$, where $I_{0k}$ is 
the ensemble-averaged intensity in channel $k$.

The mean intensity in channel $k$ is
$\bar I_k = {{1}\over{N_o}} I_{0k} \sum_{j} A(t_j)$,
and the variance from variations in $A$, the noise, is 
$\delta (I_k ^2) = {{1}\over{N_o^2}} I_{0k}^2 \sum_{j}  A(t_j)^2$.
For a constant-intensity source,
all the $A(t_j)=1$, and this expression for noise becomes the Dicke Equation, Eq. \ref{eq:dicke_eq}.
More generally, if we demand that an ensemble average over noise yield a particular observed, average spectrum,
then we require $\sum_{j} A(t_j) \equiv \alpha N_o$ for one value of $\alpha$.
It is then easy to show (for example, by the method of Lagrange multipliers) 
that among the possible sets of $A(t_j)$ in this restricted set, the minimum noise $\delta (I_k ^2)$ is obtained when all of the $A(t_j)$ are equal: $A(t_j)=\alpha$.
Thus, variation in amplitude increases self-noise.

As a simple example, relevant to pulsar observing and to pulsar gating,
suppose that the source is ``on'' for $d_P N_o$ of the samples,
and ``off'' for the remainder.
Thus, the source is pulsed, with a rectangular pulse with duty cycle $d_P$.
If we demand that the mean measured intensity be $I_{0k}$, then $A(t_j) = 1/d_P$ when the source is ``on'' and $A(t_j)=0$ otherwise.
We then find for the noise $\delta (I_k ^2) = {{1}\over{d_P N_o}} I_{0k}^2$.
This result takes the form of the Dicke equation, but with the reduced number of samples offered by the source.
If the instrumental gate is narrower than the pulse, the same result holds, but with the duty cycle of the instrument rather than the pulsar.
This example is similar to the discussions of noise for pulsars by \citet{Viv82}, and \citet{Kou01};
all assume  a pulsed source with constant on-pulse flux density (that is, rectangular pulses).

If individual pulses vary in amplitude, or the pulse amplitude varies while spectra are accumulated
within one pulse, 
then noise increases further.
In \citet{Gwi10} we present a calculation including background noise and signal of arbitrary strength;
we recover the noise polynomials, Eqs.\ \ref{eq:visibility_noise_polynomial} and\ \ref{eq:intensity_noise_polynomial},
but with an increased value of $b_2$ for a smaller duty cycle, or for amplitude variations.
From the standpoint of signal and noise alone, effects of amplitude variations can be represented as a decreased ``effective'' duty cycle.
However, the number of independent measurements may remain unchanged, despite the variation in noise.
Pulsar B0834$+$06 shows such amplitude variations strongly, as noted above;
indeed, most pulsars observed with sufficient signal-to-noise ratio exhibit such variations.
As we demonstrate below in \S\ref{sec:diffnoise} and \S\ref{sec:global_dist_vlbi},
this pulsar shows noise much greater than that expected on the basis of the Dicke equation,
even when taking pulsed emission and gating into account.
We suggest that additional amplitude variations, within and among pulses, may be responsible for the variations.

\subsubsection{Short-Term Amplitude Variations}\label{sec:short-term_amplitude_variations}

Formation of a single realization of a spectrum, of $N_{\nu}$ channels, requires $N_{\nu}$ samples of the electric field $E$.
If the signal varies within the time span of accumulation of $N_{\nu}$ samples, 
then noise is given by the Dicke equation, but is correlated among spectral channels.
This time span is
about 300\ $\mu$sec for our observations.
Such time variations have been observed directly, and statistically, for a number of pulsars 
\citep{Han71,Cog96,Jen01,Bil08} and have been proposed as a fundamental component of pulsar emission \citep{Cor76}.
Most notably, \citet{Kar78} observed variations of our program pulsar, B0834$+$06, 
with resolution of $10\ \mu{\rm sec}$ at an observing frequency of 102.5\ MHz, and 
found strong variations with a often-expressed periodicity of 160 to $700\ \mu{\rm sec}$.
In general, such short-term variations in emission from a broadband source 
do not affect the average spectrum or correlation function,
which are determined by propagation effects.
They do introduce correlations of noise among spectral channels,
and affect the distribution of noise in the lag-rate domain.
We discuss these effects in detail elsewhere \citep{Gwi10}.

As a simple example, suppose that the intrinsic emission from the pulsar consists of a spike of
electric field, nonzero only at a single time sample.
Propagation through the interstellar medium will convolve that spike with the impulse-response function,
a combination of effects of dispersion \citep{Han71} and multipath propagation \citep{wil72}.
Fourier transform of that function yields the corresponding spectrum: the scintillation spectrum.
The random amplitude and phase of the original emission spike will change the amplitude of that spectrum by a single factor.
Thus, for emission of a single spike, self-noise corresponds to a random gain factor for the entire spectrum.
Each spectral channel will be subject to noise; but, that noise will be perfectly correlated among channels.
In contrast, if the source emission is noiselike and of constant intensity, averaging will yield the same average spectrum,
but completely uncorrelated noise between channels.
Thus, in this example, the noise is identical for the spectrum, even though the number of independent samples is different.
Indeed, the noise in the spectrum follows the behavior predicted by Eq.\ \ref{eq:dicke_eq} 
\citep[see][]{Gwi10};
this is a consequence of the fact that the variance of noise in a single spectrum is equal to the square of the spectrum \citep{Bro06}. 
The noise in the lag-rate domain will be concentrated for the spike, but stationary in frequency for constant-intensity emission.

\subsection{Secondary Spectrum}\label{sec:FT_convention}

\subsubsection{Transform to the Lag-Rate Domain}

We measure spectra, $I(\nu,t)$, as a function of frequency $\nu$, and index them by the time of observation $t$ to create the dynamic spectrum $I(\nu,t)$.
The 2D Fourier transform of $I(\nu,t)$ is the correlation function in the lag-rate domain, $\tilde I(\tau,\omega)$, where $\tau$ is the variable conjugate to $\nu$ and 
$\omega$ is conjugate to $t$.
We denote $\tau$ as ``delay" and $\omega$ as ``rate", in agreement with the standard notation in
interferometry \citep{tms01}.
For interferometric data, $V(\nu,t)$ is the dynamic cross-power spectrum, 
and $\tilde V(\tau,\omega)$ is its 2D Fourier transform, the correlation function in the lag-rate domain.
Its square modulus is the secondary spectrum.
Note that $I$ must be purely real,
whereas $\tilde I$, $V$, and $\tilde V$ may be complex.

We choose the convention for Fourier transform so that
a signal of unit amplitude, constant in time and frequency,
appears in the lag-rate domain as a spike of amplitude 1, at the origin \citep[see][\S 2.2.2]{Gwi06}.
Mathematically, we define
\begin{equation}
\tilde V(\tau,\omega) = {{1}\over{N_{2}}} \sum_{\nu, t} \exp\left\{i 2 \pi (\tau\nu+\omega t)\right\} V(\nu, t) ,
\label{eq:FT_convention}
\end{equation}
and likewise for $\tilde I$ and $I$.
Here, $N_{2} = N_{\nu}\times N_{t}$ is the number of samples in the dynamic spectrum,
equal to the product of frequency channels $N_{\nu}$ and number of spectra gathered $N_t$.
The secondary spectrum is the square modulus of the lag-rate correlation function: $\tilde I \tilde I^*$ or $\tilde V \tilde V^*$.

\subsubsection{Parseval's Theorem and Noise}\label{sec:parseval_noise}

Parseval's theorem relates the mean squared intensity (or visibility) in the dynamic spectrum with that in the secondary spectrum.
Indeed, this theorem holds for any Fourier transform pair of functions.
For our convention, Eq.\ \ref{eq:FT_convention},
\begin{equation}
\sum_{\tau,\omega} |\tilde I(\tau,\omega)|^2 = {{1}\over{N_{2}}} \sum_{\nu, t} | I(\nu, t)|^2 ,
\label{eq:parseval}
\end{equation}
where the sums run over all samples.
Parseval's theorem is related to conservation of energy in physics;
however, in our situation application is to the square of intensity or visibility, and thus roughly to the square of power.

Parseval's Theorem also holds for noise;
as it does for the observed and ensemble-average intensity or visibility.
Noise in the dynamic spectrum is
$\delta V(\nu, t) = V(\nu, t) - \langle V(\nu, t)\rangle_n$,
whereas noise in the lag-rate domain is
$\delta \tilde V(\tau,\omega) = \tilde V(\tau,\omega) - \langle \tilde V(\tau,\omega)\rangle_n$.
The Fourier transform is linear,
so $\delta \tilde V$ is the Fourier transform of $\delta V$.
Thus,
\begin{equation}
\sum_{\tau,\omega} |\delta \tilde V(\tau,\omega)|^2 = {{1}\over{N_{s}}} \sum_{\nu, t} |\delta V(\nu, t)|^2 .
\label{eq:parseval_for_noise}
\end{equation}
We make use of this expression in \S\ref{sec:noise_box} below.

\subsubsection{Correlated and Uncorrelated Noise}

If noise is uncorrelated among samples of the dynamic spectrum, then it is stationary in the secondary spectrum.
Mathematically, if the noise is uncorrelated, then:
\begin{equation}
\langle \delta V(\nu_0,t_0) \; \delta V^*(\nu_1,t_1)\rangle_n = 0\quad {\rm for\ }\nu_1\neq \nu_2{\rm\ or\ }t_1\neq t_2 .
\label{eq:uncorrelated_noise}
\end{equation}
This situation holds, for example, 
when the source emits a noiselike electric field, with constant intensity, over the period over which a spectrum is accumulated \citep{Ric75}. 
A constant-intensity source, modulated by a spectral and time envelope, as from scintillation, as described in \S\ref{sec:short-term_amplitude_variations},
will also produce uncorrelated noise. 

Noise in the Fourier-conjugate domain of the lag-rate autocorrelation function is the Fourier transform of noise in the dynamic spectrum. 
The Fourier transform of white noise is white noise \citep{Pap91}.
By the convolution theorem, the Fourier transform of the modulated spectrum is the convolution of white noise with the Fourier transform of the 
spectral and time envelope, as given by Eq. \ref{eq:visibility_noise_polynomial} or \ref{eq:intensity_noise_polynomial} and the average intensity.
The result of this convolution is stationary with $\tau$ and $\omega$; thus, uncorrelated noise is stationary in the lag-rate domain, and in the secondary spectrum.

However, if noise is correlated between samples in the original dynamic spectrum $I(\nu, t)$,
then it need not be stationary in the lag-rate domain of $\tilde I(\tau,\omega)$.
As an example, consider the above example of a single spike of emission, resulting in perfectly correlated noise over a spectrum.
The Fourier transform of the average spectrum will be the average correlation function.
For spectral variation resulting from interstellar propagation as discussed in \S\ref{sec:short-term_amplitude_variations} above, 
this will be the autocorrelation of the impulse-response function.
Noise will then simply be a scaling of that average by a different factor for each spectrum. 
Thus, the distribution of noise will simply be that of the average correlation function,
times some factor.  It certainly need not be distributed evenly over the spectrum; indeed, typically it will be peaked near the origin.
We defer detailed discussion to a more complete mathematical treatment \citep{Gwi10}.

\subsection{PSR B0834$+$06}\label{sec:PSR0834}

PSR\ B0834$+$06 has a period of 1.27\ sec \citep{Tay00}. 
Its intrinsic duty cycle is $d_P=1.8\%$ at observing wavelengths of $\lambda \approx 1$\ m. This duty cycle includes 90\% of mean flux density,
as determined from our Puschino observations.
The pulsar shows frequent nulls, or absence of pulses, and 
has a modulation index of close to 1;
particularly strong modulation appears with a period of 2.17 pulsar periods \citep{Tay75,Rit76,Kar86,ad05,ran07}.
Modulation persists on timescales from an entire pulse to $<160\ \mu{\rm sec}$ \citep{Kar78}.
It has a dispersion measure of $DM=12.86\ {\rm cm}^{-3}{\rm pc}$, and
lies at a distance of about $D=0.72\ {\rm kpc}$, 
as determined from
a model for the interstellar medium \citep{Cor02}.
As discussed in \S\ref{scint_params}, we adopt a value of $\Delta \nu_d = 0.57$\ MHz
for the scintillation bandwidth of the pulsar
at our observing frequency.
For a uniform distribution of scattering material,
we then expect typical angular scattering
of $\theta_H=1.0$\ mas \citep{Gwi93}.
This is the diameter of an assumed 
circular-Gaussian ``scattering disk'' seen from the Earth.

PSR B0834$+$06 commonly shows a scintillation arc in the secondary spectrum
\citep{Hil03, Wal08}. 
We observe this as well, for both instruments, 
as discussed in \S\ref{sec:compare_secondary} below.
Using a software correlator with 131072 spectral channels, \citet{Bri10} found that the arc extends to
delays of milliseconds, well outside the range of our secondary spectrum.
We also see a more isolated feature in the secondary spectrum, which we call the ``clump''.
Such features are not uncommon \citep{Hil03}. 

\subsection{Note on Units and Calibration}\label{sec:flux_dilution}

\subsubsection{Units for Flux Density}\label{sec:flux_density_units}

In this paper, we report flux densities in Jy,
diluted over the full period of the pulsar, as is traditional;
or equivalent instrumental units (VLBA correlator units or ``V.c.u.'', and WAPP units or ``W.u.'').
For example, if the pulsar had a rectangular pulse with duty cycle $d_P$,
we would report a flux density $I_S$ if the peak flux density is $I_S/d_P$.
This is most easily incorporated into the noise calculations above
as a rescaling of the flux-density scale.
In principle, Fourier transforms affect units.
Here, the Fourier-transform variables to the lag-rate domain are time and frequency, so dimensionally the units for flux density are the same in both domains,
although the quantities are different.

\subsubsection{Effects of Calibration}

We perform calibration in 3 successively larger spheres:
calibration internal to observations with a single instrument,
inter-calibration of instruments,
and calibration in Janskys averaged over pulse phase.
Only the first of these calibrations is important to our results for noise.
The other 2 allow useful comparisons, but are not important to our conclusions.

Internal calibration includes zeroing the average phase of the visibility for each time step,
as discussed in \S\ref{sec:phase_ref}.
Fitting for the offset of the intensity distribution (\S\ref{sec:global_dist_sd})
and measurement of the amplitude variations (\S\ref{sec:amplitude_variations})
can also be placed in this category;
these are included as parameters for the estimates of noise.

We perform inter-calibration between VLBA and WAPP by comparing flux density in
the overlapping portion of the spectrum,
for identical times (\S \ref{sec:amplitude_variations}).
The accuracy of this calibration is likely about 10\%, the approximate difference of results from proportionality.
The major discrepancy between the two instruments is likely the difference in width of their pulsar gates:
the much narrower WAPP gate does not always capture the entire pulse.  

We calibrate to Jy by comparing background noise for the Arecibo-Jodrell baseline
with the value expected from telescope parameters (\S\ref{sec:diffnoise_vlbi});
results are in good agreement with fits for the mean flux density of the pulsar (\S\ref{sec:global_dist})
and with tabulated values \citep{lor95}.
This calibration is probably good to a factor of 2.  The calibration lends confidence to the notion that
our observations are indeed detecting the physical effects that we model.

\section{OBSERVATIONS}\label{sec:observations}

\subsection{Measurement of Scintillation Parameters}

\subsubsection{Puschino Observations}

In order to re-determine the scintillation bandwidth and timescale for PSR\ B0834+06, we observed the pulsar at a frequency of 111.07 MHz using the high-sensitivity BSA telescope of the Pushchino Radioastronomy Observatory, of the Lebedev Institute of Physics. The observations took place in October 2001. The 128 channels of 
the receiver, with a bandwidth of 1.25 kHz per channel, were recorded every 1.23 ms in a pulse gate of  300 ms duration synchronized with the pulsar period.
The total observing time for one observation was 3.2 min. The instrument and observing technique are described in more detail by \citet{Mal95}. 
To obtain dynamic spectra, we averaged the signal over the fraction of  pulse phase with amplitude $\geq 0.2$ times the average amplitude of the pulse, and then summed the signal over 5 pulsar periods (6.37 sec).

\subsubsection{Scintillation Parameters}\label{scint_params}

We define the scintillation timescale $t_d$ as the time lag where the normalized cross-correlation coefficient of spectra separated in time falls to $1/e$. We define the scintillation bandwidth of $\Delta \nu_d$ as half-width at half-maximum of the mean cross-correlation function,
with frequency, of adjacent pulses.
From our Puschino observations, 
we determined a scintillation timescale of $t_d = 84\pm 20$\ sec and a scintillation bandwidth of $\Delta \nu_d = 3.5\pm 0.5$\ kHz. The quoted errors are the standard deviations of these parameters, over 10 days of observations. 

\begin{figure}[t]
\epsscale{0.80}
\plotone{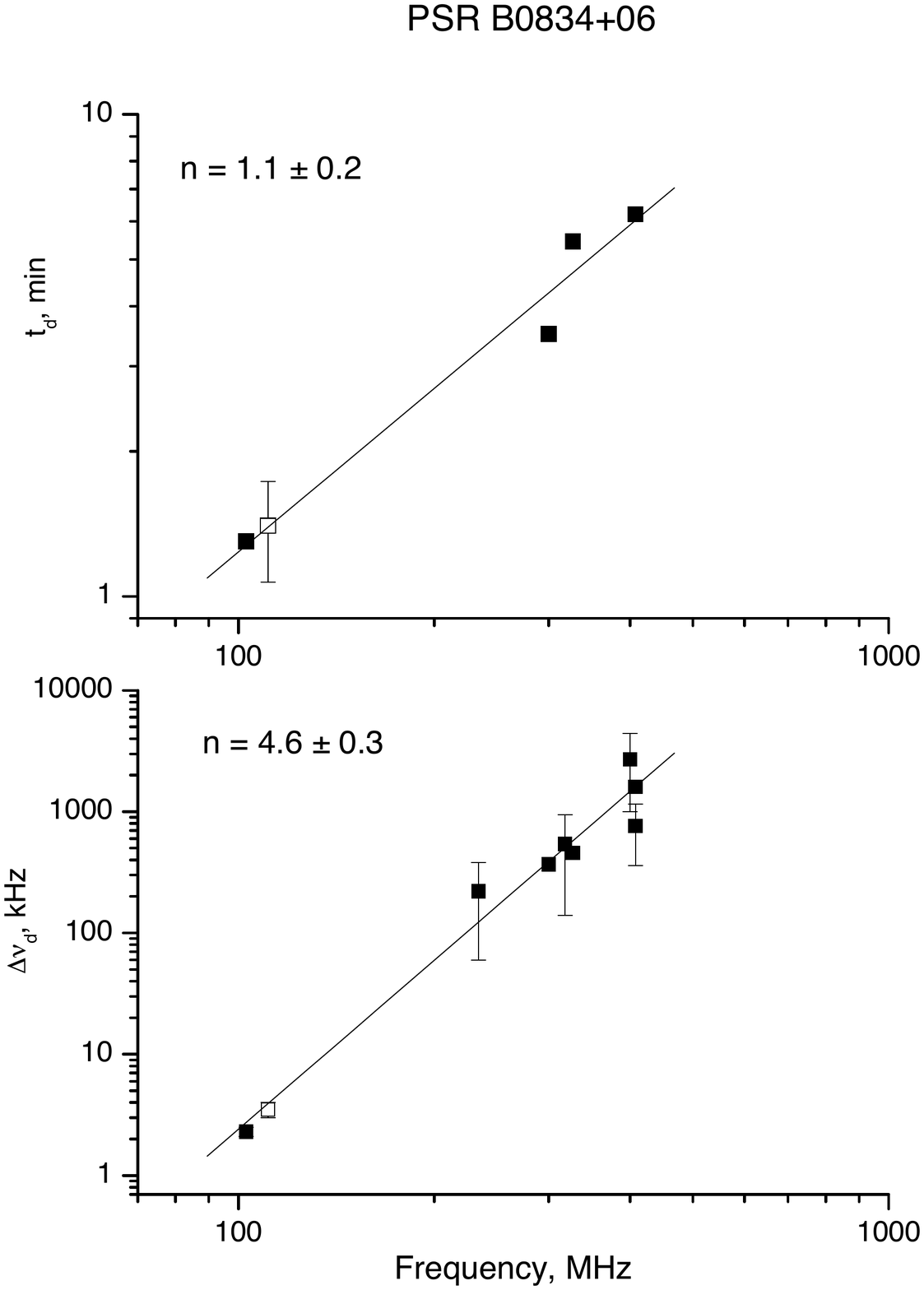}
\figcaption[]{The dependence of diffractive scattering parameters on the observing frequency. Straight lines correspond to fits to the data points.
\label{fig:Tania_Fig2}}
\end{figure}

Figure\ \ref{fig:Tania_Fig2} presents the dependence of the diffractive parameters on the observing frequency $\nu$,
as determined from our observations and the literature. 
Here we use our data (open squares) and  published data at 102.7 MHz \citep{Smi92,Mal95}, 234 MHz \citep{Hug69}, 400 MHz \citep{Hug69}, 300 MHz \citep{Bal85},  327 MHz \citep{Bha98}, 408 MHz \citep{Lan71,Smi85}. 
We converted the data of \citet{Bal85} to our definition of $\Delta \nu_d$ using the coefficient 0.3. Linear fits to the log-log scale result in $t_d \propto \nu^{1.1}$ and $\Delta \nu_d \propto \nu^{4.6}$. These indices agree with those predicted for a Kolmogorov model within the errors. Using this linear fit we found a scintillation bandwidth of $\Delta \nu_d = 0.57$\ MHz and a scintillation timescale of  
$t_d = 290$ sec, at the $\nu=327$ MHz observing frequency of the observations discussed  below.

\subsection{Single-Dish and VLBI Co-Observations}

We observed a number of pulsars, including PSR\ B0834+06, on 2004 Oct 22.  
We used the 25-m dishes of the 
Very Long Baseline Array
and the European VLBI Network, including the 76-m Lovell Telescope at Jodrell Bank and the 305-m antenna at 
Arecibo Observatory.
We observed near $\nu = 327$ MHz, where the arcs tend to be common.
All pulsars were detected, but scintillation substructure and arcs were seen most clearly for PSR\ B0834+06,
observed from 10:04 UT to 11:18 UT.
Figure\ \ref{fig:align} shows the resulting dynamic spectra from the two instruments, aligned in frequency.
We focus on this pulsar, and this time interval, in this paper.

\subsubsection{VLBA}

We cross-correlated data from different antennas using the VLBA correlator.
We correlated both circular polarization states and averaged results to find Stokes $I$.
We obtained 1024 channels with bandwidths of 7.8125 kHz,
sampled every 10 sec and gated synchronously with the pulsar pulse.
The duty cycle of the gate for the VLBA was $d_V=30\%$.
Because the VLBA correlator used the same engines for pulsar gating and for spectroscopy,
hardware limits prevent simultaneous high spectral resolution, and a narrow pulsar gate.
The spectra spanned 319.99 to 327.99\ MHz, an 8\ MHz bandwidth.
The VLB observations covered the same span in UT as the single-dish observations, 
but were periodically broken for 90\ sec to reverse tape direction. 
A few channels showed interference,
visible as increased, variable visibility in particular channels.  Interference does not correlate
between antennas, but can drastically raise the
background noise in the affected channels.

\subsubsection{Arecibo: WAPP}

At Arecibo observatory,
we formed single-antenna dynamic spectra using the Wide-Band Arecibo Pulsar Processor,
or WAPP \citep{Dow00}.
We obtained spectra of 2048 channels with bandwidths of 3.052\ kHz, sampled every 10 sec.
The spectra spanned the frequency range of $\nu=319.37$ to 325.62\ MHz, a bandwidth of $\Delta\nu=6.25$\  MHz. 
Thus, accumulation of one spectrum required $2048/(6.25\times 10^6\ {\rm MHz})=327.68\ \mu{\rm sec}$.
The spectra were gated synchronously with the pulsar pulse,
and dedispersed using the incoherent method described by \citet{Vou02}, 
to improve signal-to-noise ratio.
The duty cycle of the WAPP gate was $d_{W}\approx 1\%$, near the peak of the pulse. 
We observed for 4400 sec, on 2004 October 22, between 08:00 UT and 
09:15 UT.
Thus, during the 10-sec integration time to form one spectrum,
we averaged approximately $10\ {\rm sec}\times 0.01/327.68\ \mu{\rm sec}= 305$ spectra together,
depending on the relative phase of pulsar pulse and integration window.

\subsection{Calibration}

In order to secure the generality of our conclusions, 
in the face of our non-standard observing modes and surprising results for the noise,
we sought to calibrate the data using the fewest 
and most basic assumptions possible.
Thus, although automatic calibration to Jy is available and effective, particularly for the VLBA,
we elected not to use this path; although
we did make the ``Tsys'' corrections for the VLBA,
which correct for the continual adjustment of analog gains 
to keep the 2-bit samplers at optimal settings.
Similarly, we did not correct for amplitude variations across the spectral passband, or truncate the
edges of the passband (where gain is presumably lower), for either VLBA or WAPP.  
Analysis with these corrections implemented did not affect our results for noise or signal,
in the dynamic spectrum or the lag-rate correlation function. We elected to present the uncorrected results here.

\subsubsection{Phase Calibration of VLBA Data}\label{sec:phase_ref}

A variety of effects produce phase offsets of VLBI data;
ionospheric propagation is probably the largest for our observations \citep[see][]{tms01}.
Additional effects include propagation in the neutral atmosphere, an optical path length of
about 2\ m, or about $4\pi$ radians, at zenith.
Clock errors or source and station position errors are expected to be a small fraction of $2\pi$ radians, and to change only slowly with time, 
for our strong, often-observed sources at these well-calibrated antennas.
Errors in estimated delay introduce a slope of phase with frequency;
we expect that these are far less than one lag (125\ nsec),
corresponding to a slope of $<< 2 \pi$ radians across our observing band.
Indeed, we observe no significant variation of phase across the band.

We removed the variation of phase with time for the program pulsar.
For each spectrum, we found the average phase, and then 
rotated all data to subtract that phase.  This is 
equivalent to phase-referencing to an intensity-weighted average phase,
over the sky image. The phase corrections were typically less than one radian, rising to $\approx 2$\ radians at the most, at the
beginning of the observations; they were constant with frequency to the accuracy of our observations.

\subsubsection{Inter-Calibration of VLBA and WAPP Amplitudes}\label{sec:inter-calibration}

\begin{figure}[t]
\epsscale{1.00}
\plotone{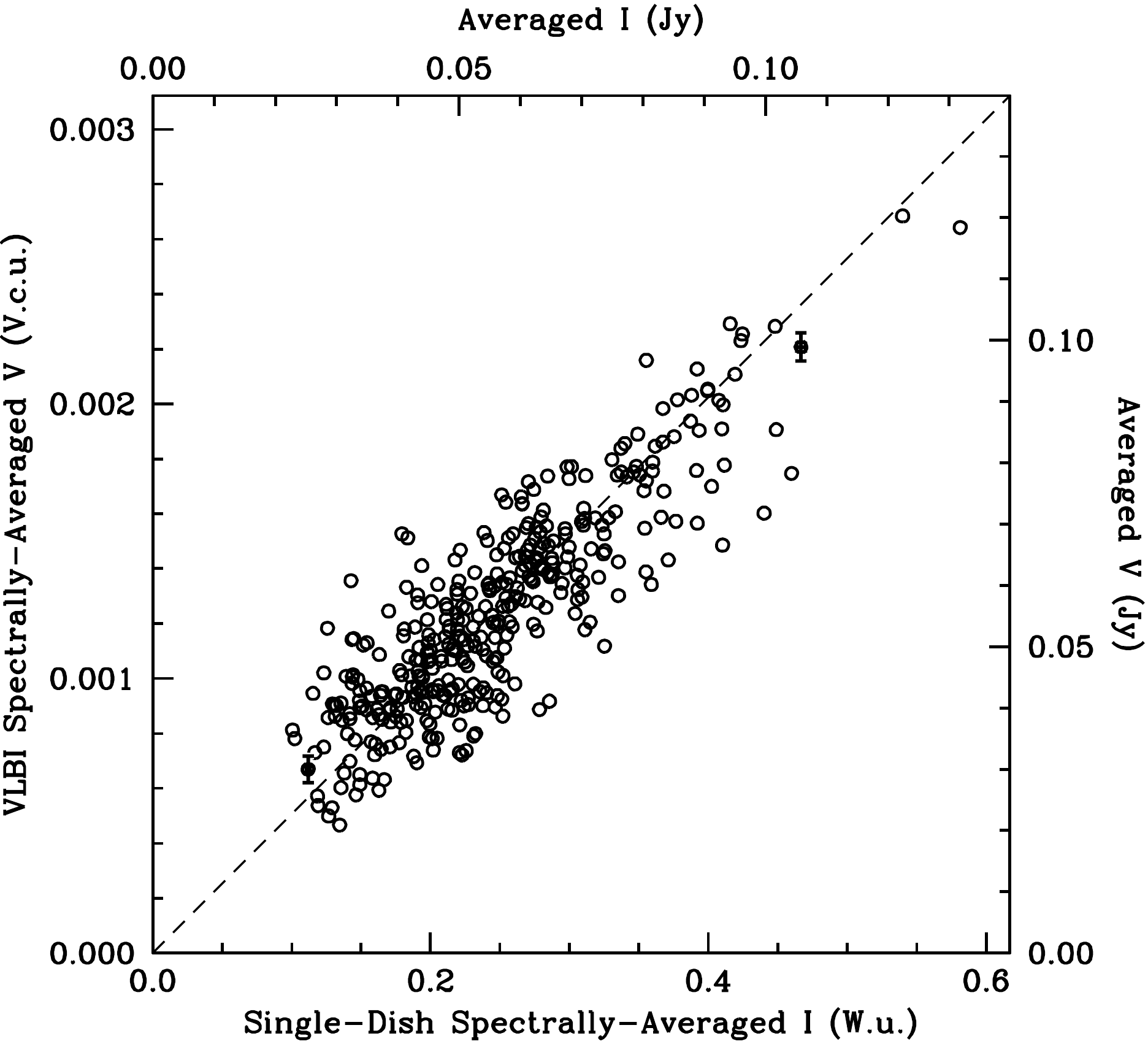}
\figcaption[]{Comparison of the portions of 
dynamic spectra that overlap in frequency, 
showing average real part of the interferometric visibility on the 
Arecibo-Jodrell baseline (vertical)
plotted with average intensity in the Arecibo WAPP single-dish spectrum (horizontal).
Broken line shows the best-fitting relationship,
used for relative calibration of the data.
Top scale and left scale show inferred Jy, averaged over the spectrum at each time,
and over the pulsar pulse.
Error bars show uncertainty resulting from noise,
for 2 typical points.
\label{fig:compare_sums}}
\end{figure}

We calibrate the WAPP observations by comparison with the VLBA observations.
As discussed in \S\ref{sec:amplitude_variations} below,
scintillation and pulse-to-pulse variability
cause large variations in the flux density of the source.
We compare the two, in the spectral range of overlap of these instruments, to find the relative amplitude
measured by the two instruments.
We compared the real
part of the WAPP spectrum with the real part of the VLBA spectra,
as Figure\ \ref{fig:compare_sums} shows.
(Of course, the symmetry of the autocorrelation function sets the imaginary part 
of the WAPP spectrum to zero, and the imaginary part of the VLBA spectrum appears to contain only noise).
For each time interval, we averaged the spectra over the range in which they overlap.
We cross-correlated these averages, 
and chose the peak of the correlation function as the time offset that aligned the two series.
The ratio of the two measurements then yields the relative calibration.
The best-fit ratio is 
197\ V.c.u.=1\ W.u.
Here we are comparing signal with signal, within the instrumental gates, but diluted over the entire pulse period as discussed in \S\ref{sec:flux_dilution} above.
The inter-calibration of VLBA and WAPP is likely accurate to about 12\%, the variation of the points in Figure\ \ref{fig:compare_sums} from proportionality.
Note that calibration 
does not affect $b_2$, which is dimensionless.
Nor does it affect the relation of noise in the dynamic and secondary spectra, as given by Parseval's Theorem.

\subsubsection{Amplitude Calibration to Jy}\label{sec:calibration_to_Jy}

We calibrate the data in Jy to provide scale for sizes of quantities, in the context of the discussion of noise in 
\S\ref{sec:flux_dilution} above.
We estimate the noise for VLBA observations from the 
level of background noise,
$b_0=1.52\times 10^{-6}\ {\rm V.c.u.}^2$, estimated as described in \S\ref{sec:diffnoise_vlbi} below.
We compare this measured value with the expected value, given by Eqs.\ \ref{eq:dicke_eq} and \ref{eq:visibility_noise_polynomial} for noise only, $I=I_n$. 
The expected noise is the product of the
system-equivalent flux densities of the two stations,
$I_{nA}=12$\ Jy at Arecibo and $I_{nJ}=132$\ Jy at Jodrell Bank,
scaled by the instrumental duty cycle $d_V$ and
divided by the number of samples $N_o$.
Thus, in a bandwidth of 7.8125\ kHz and an integration time of 10\ sec, averaging two polarizations,
and with a duty cycle of $d_V=30\%$ for the pulse gate, we expect noise of 
$b_0 = 0.30^2\cdot (132\ {\rm Jy} \cdot 12\ {\rm Jy})/(2\cdot 7812.5\cdot 10\cdot 0.3) = 3.0\times 10^{-3}\ {\rm Jy}^2$
for real and imaginary components in each channel, in the absence of signal (and consequently, of self-noise).
Note that here, the instrumental duty cycle appears twice in the numerator,
to account for expressing the pulsar flux density as averaged over an entire pulse period as discussed in \S\ref{sec:flux_dilution} above,
and once in the denominator to reflect the smaller number of samples.
By equating to our observed $b_0$,
$1\ {\rm V.c.u.} = 63\ {\rm Jy}$.  
From this, and the calibration of V.c.u., we find 
$1\ {\rm W.u.} =0.32\ {\rm Jy}$.
Note again that this calibration does not affect our inferred value for $b_2$ or with calculation of noise in the secondary spectrum
via Parseval's Theorem.
Thus, although calibration to Jy provides the benefit of placing VLBA and WAPP on the same footing,
it is tangential to our arguments.

\section{ANALYSIS}\label{sec:analysis}

\subsection{Dynamic Spectra: Frequency-Time Domain}\label{sec:dynspec}

The dynamic spectrum of pulsar B0834$+$06 shows scintillation, and substructure.
Figure\ \ref{fig:align} compares 
dynamic spectra for the VLBA and Arecibo.  The spectra have been shifted horizontally so that they
have the same frequency scale, and overlapping portions of the spectral range match. 
The spectra are clearly quite similar in the overlapping range.
The WAPP attains much lower noise (because of the WAPP's narrower pulse gate,
and the greater collecting area of Arecibo than Jodrell Bank).
The scintillation maxima have dimensions of about the expected size:
about $0.57\ {\rm MHz}\times 290\ {\rm sec}$.
Finer structure is visible, most clearly in the single-dish spectrum,
as a somewhat ``diagonally striped'' appearance of the pattern, mostly from lower left to upper right,
although structure with a variety of scales and axes is clearly present.  
The finer parts of this spectrum contribute to the scintillation arc and the clump in the secondary spectrum.
We also see intrinsic variations in the flux density of the pulsar;
these produce horizontal stripes in Figure\ \ref{fig:align}.

\subsection{Intrinsic Amplitude Variations}\label{sec:amplitude_variations}

The WAPP and VLBA 
yield spectra over different frequency ranges, as Figure\ \ref{fig:align} shows.
The overlapping spectral range includes 720 channels for the VLBI data, 
and 1846 channels for the single-dish data.
We aligned these in time by cross-correlating spectra from the two instruments, summed over the overlapping range.
The peak of the correlation was quite sharp:
it loses about 11\% of its total range of variation, at a lag of only $\pm 1$ time sample from the peak.
If scintillation is responsible for most of the range of variation, and intrinsic variation is about 12\% as estimated from the WAPP data,
we would expect about this peak height.

Figure\ \ref{fig:compare_sums} shows the comparison, for the averages over the overlapping spectral range.
The best-fit ratio yields the relative calibration of VLBA and WAPP measurements, discussed in \S \ref{sec:inter-calibration} above.
Both intrinsic variations and scintillation should produce correlated variations of VLBA and WAPP, parallel to the dashed line.
The relative variation produces variations perpendicular to the dashed line.
For comparison, we show the expected 1-standard-deviation error bars for a couple of points in the figure,
as found from using noise from time-differencing
in \S\ref{sec:diffnoise_vlbi} and \S\ref{sec:diffnoise_sd} below.
These error bars reflect the effects of self-noise, evident from the change the horizontal, single-dish direction;
and the expected reduction in noise by a factor of $1/\sqrt{N_s}$, by averaging over $N_s$ spectral channels.
Clearly, the relative variation is greater than expected from noise alone.

\begin{figure}[t]
\epsscale{0.80}
\plotone{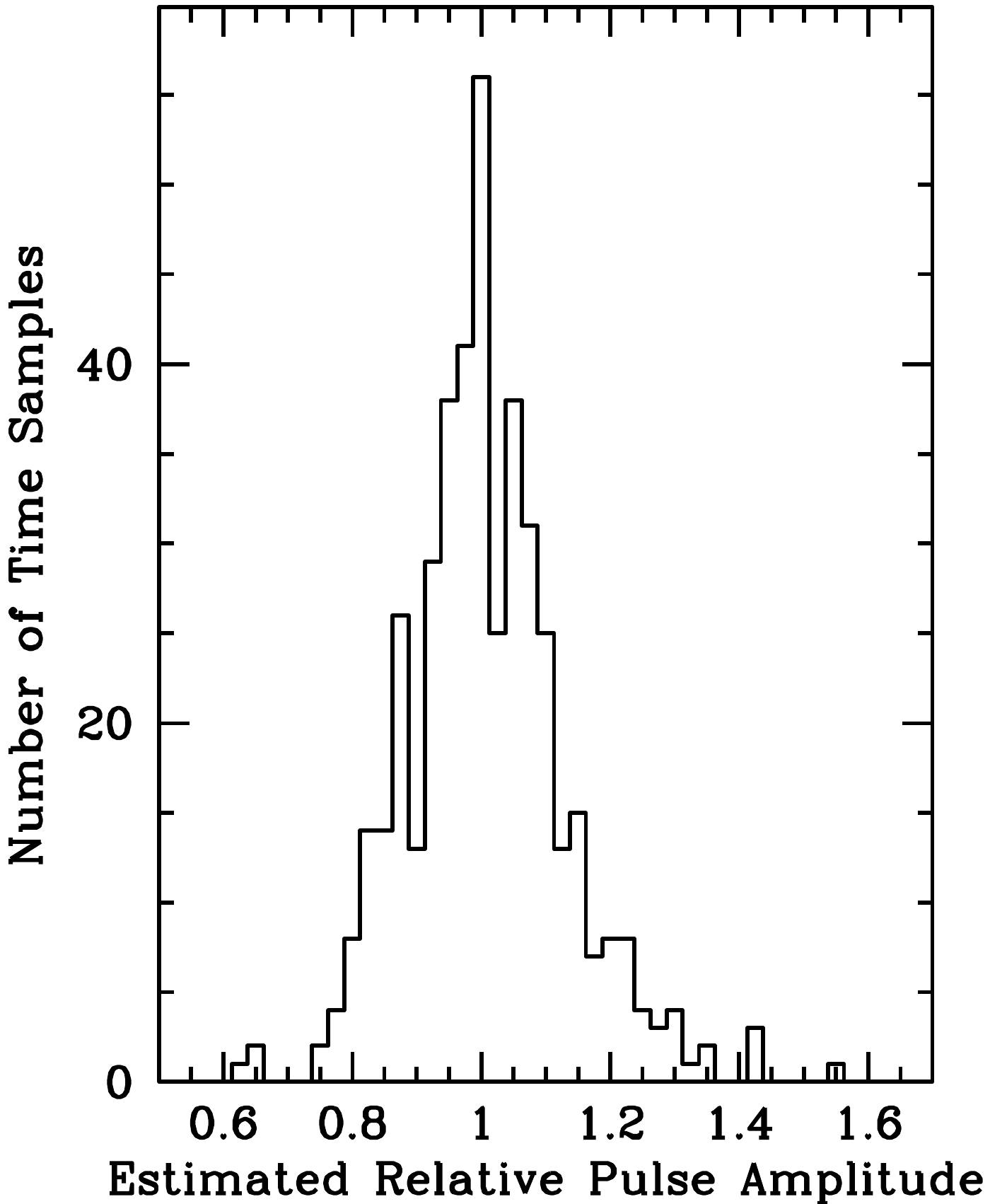}
\figcaption[]{
Variations in intrinsic flux density of the pulsar,
estimated as the square root of the ratio of successive WAPP spectra, summed over frequency.
The standard deviation is 12\%.
\label{fig:sumt}}
\end{figure}

Between successive integrations, 
the scintillation spectrum will vary little,
but intrinsic variations can be significant.
We can estimate these variations by comparing successive spectra.
We estimate these as the ratio
of successive spectrally-averaged amplitudes of the WAPP,
$\bar I(t_i)/\bar I(t_{i-1})$.
Because spectra are integrated over 10\ sec,
and pulse period is 1.27\ sec, each integration will comprise about 8\ pulses,
reducing the variability from the $\approx 100\%$ modulation of individual pulses.
Under the assumptions that intrinsic variations are uncorrelated between successive spectra,
and are not large,
we can estimate the variations as the square root of this ratio: $(\delta A/A)_{i} = \sqrt{ \bar I(t_i)/\bar I(t_{i-1}) }$.
Figure\ \ref{fig:sumt} shows the results for WAPP data.
The variations roughly follow a Gaussian distribution. They have standard deviation of 12\%.
Presumably averaging over pulse phase and over $\sim 8$ pulses reduces the strong variability of individual pulses to this value.

Because the WAPP pulsar gate is narrower than the pulse, 
pulse-to-pulse variations in pulse shape might
cause differences in flux density from the VLBA correlator.
Differences of tens of percent in the integrated flux density of individual pulses,
as measured in the two gates,
can easily lead to the observed differences of about 10\%.
We suggest that this is the most likely cause of the variations around perfect proportionality
seen in Figure\ \ref{fig:compare_sums}.

\subsection{Noise from Time Differences}\label{sec:diffnoise}

\subsubsection{Noise from Differences of the VLBI Dynamic Cross-Power Spectrum}\label{sec:diffnoise_vlbi}

\begin{figure}[t]
\epsscale{0.80}
\plotone{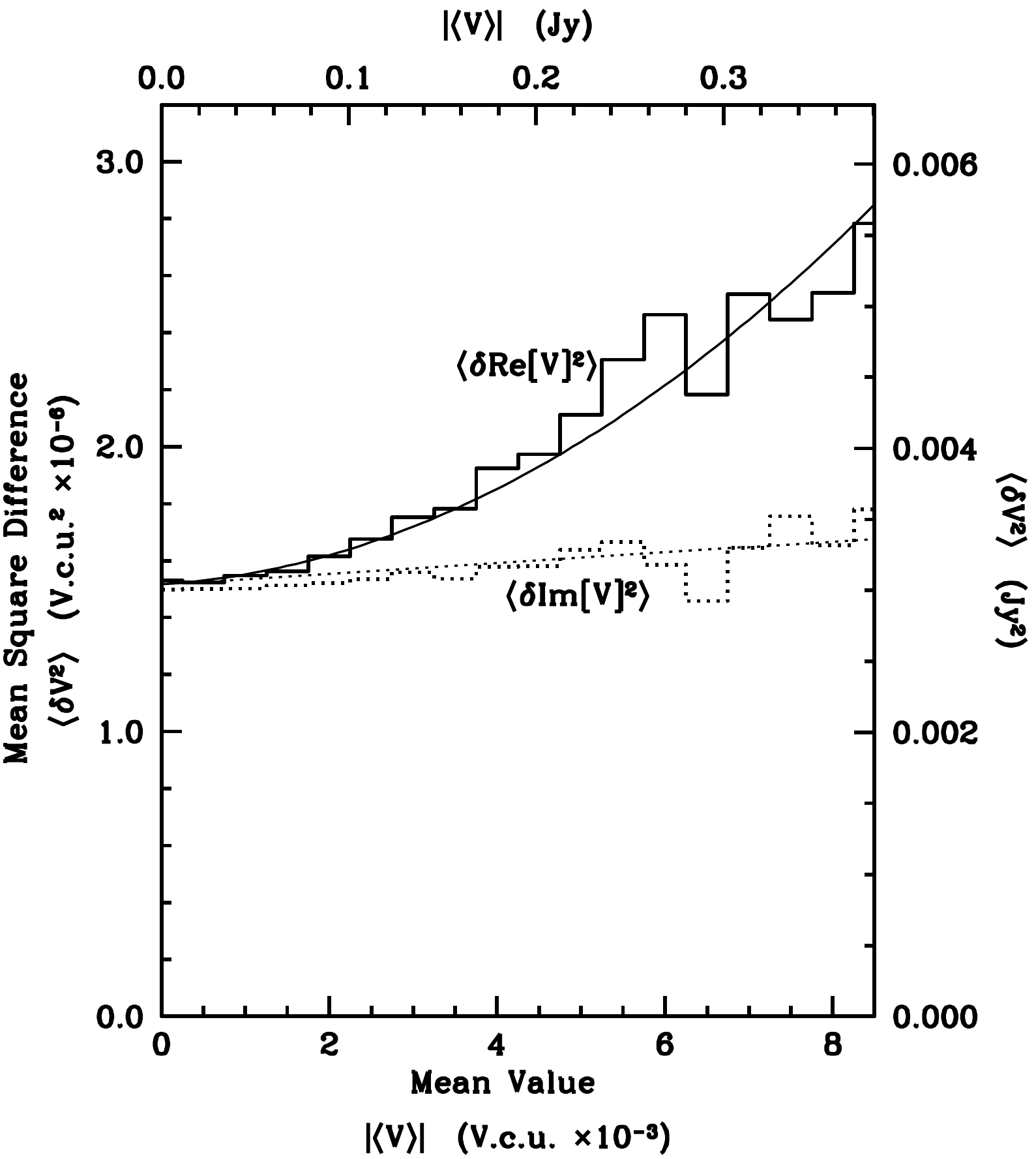}
\figcaption[]{Noise estimated by differencing  consecutive samples of visibility,
plotted with mean visibility estimated from their sum.
Upper histogram shows differences of real part, lower differences of imaginary part.
Smooth curves show best-fitting curves of the form of Eq.\ \ref{eq:visibility_noise_polynomial}.
\label{fit_vlbi}}
\end{figure}

We estimate the noise, including self-noise, by differencing consecutive samples in the same spectral channel.
Thus, differences between consecutive samples (normalized by $1/\sqrt{2}$) 
are an estimate of noise,
whereas their averages are an estimate of signal.
Noise depends upon signal; the differencing technique yields estimated 
noise as a function of estimated signal.
Figure\ \ref{fit_vlbi} shows the results.  
The noise at zero visibility, the $y$-intercept,
is the contribution from system and sky noise;
this is nearly equal for real and imaginary parts, as expected from \S\ref{sec:noise_polynomial}.

For an unresolved source, at the phase center,
we expect the real part of noise to increase quadratically with visibility,
and the imaginary part by the same constant and linear coefficients 
(see Eq.\ \ref{eq:visibility_noise_polynomial} above).
The smooth lines in Figure\ \ref{fit_vlbi} show a fit of this form to estimates from differencing.
Indeed, the observed differences follow the expected form.
We find, in units of V.c.u:  $b_0=1.52\times 10^{-6}\ {\rm V.c.u.}^2$, $b_1=1.88\times 10^{-5}\ {\rm V.c.u.}$, and $b_2=1.63\times 10^{-2}$.
We fit the data in Figure\ \ref{fit_vlbi} in the range $0<|V|<8.5\times 10^{-3}$\ V.c.u. to obtain these values.
Here, ``V.c.u.'' are VLBA correlator units.
Table \ref{parameters_table} summarizes these and other fitted parameters.

From the figure, and the signal-to-noise ratio, we expect that our fit provides a reasonable estimate of $b_0$,
and less accurate estimates of $b_1$ and $b_2$.
We use the fitted value of $b_0$ to determine the calibration of V.c.u. in Jy, as discussed in \S\ref{sec:calibration_to_Jy} above.
Samples at large amplitudes are important for estimates of $b_1$ and $b_2$, but 
the number of samples available is small.
In units of Jy, using the calibration discussed above,
we find $b_0=3.0\times 10^{-3}\ {\rm Jy}^2$, $b_1 = 8.4\times 10^{-4}\ {\rm Jy}$, and $b_2=1.63\times 10^{-2}$.

For this analysis, the parameter $b_2$ describes amplitude variations as well as noise.
It represents a difference between sequential values, proportional to the mean intensity, averaged over amplitude parameter $A$ as well as noise.
Because amplitude variations and noise variations are uncorrelated, the effects add in quadrature.
If we correct the value of $b_2$ for the longer-term amplitude variation found in \S\ref{sec:amplitude_variations}  for the WAPP, 
then we infer for the self-noise parameter $b_2 - (\delta A/A)^2 = 1.9\times 10^{-3}$.
Note that $b_2$ is dimensionless, so it is unaffected by any calibration: it takes the same value in
V.c.u. or Jy, as it would in any other system of units.

If the source had constant amplitude during integration of a spectrum, 
then from the parameters of our observation, with $d_P\approx 1.8\%$ and $d_V=0.3$, 
we would expect
$b_2 = ({{1}\over{d_P N_o}}) = 1/( 0.018 \times 7.8125 {\rm\ kHz}\times 10 {\rm\ sec})=7\times 10^{-4}$ from the discussion in 
\S\ref{sec:intermediate_term_amplitude_variations}.
Our inferred value is about 2.7 times larger, indicating that intermediate-term amplitude variations are present,
as discussed in \S\ref{sec:intermediate_term_amplitude_variations}.

\subsubsection{Noise from Differences of the Single-Dish Dynamic Spectrum}\label{sec:diffnoise_sd}

\begin{figure}[t]
\epsscale{0.80}
\plotone{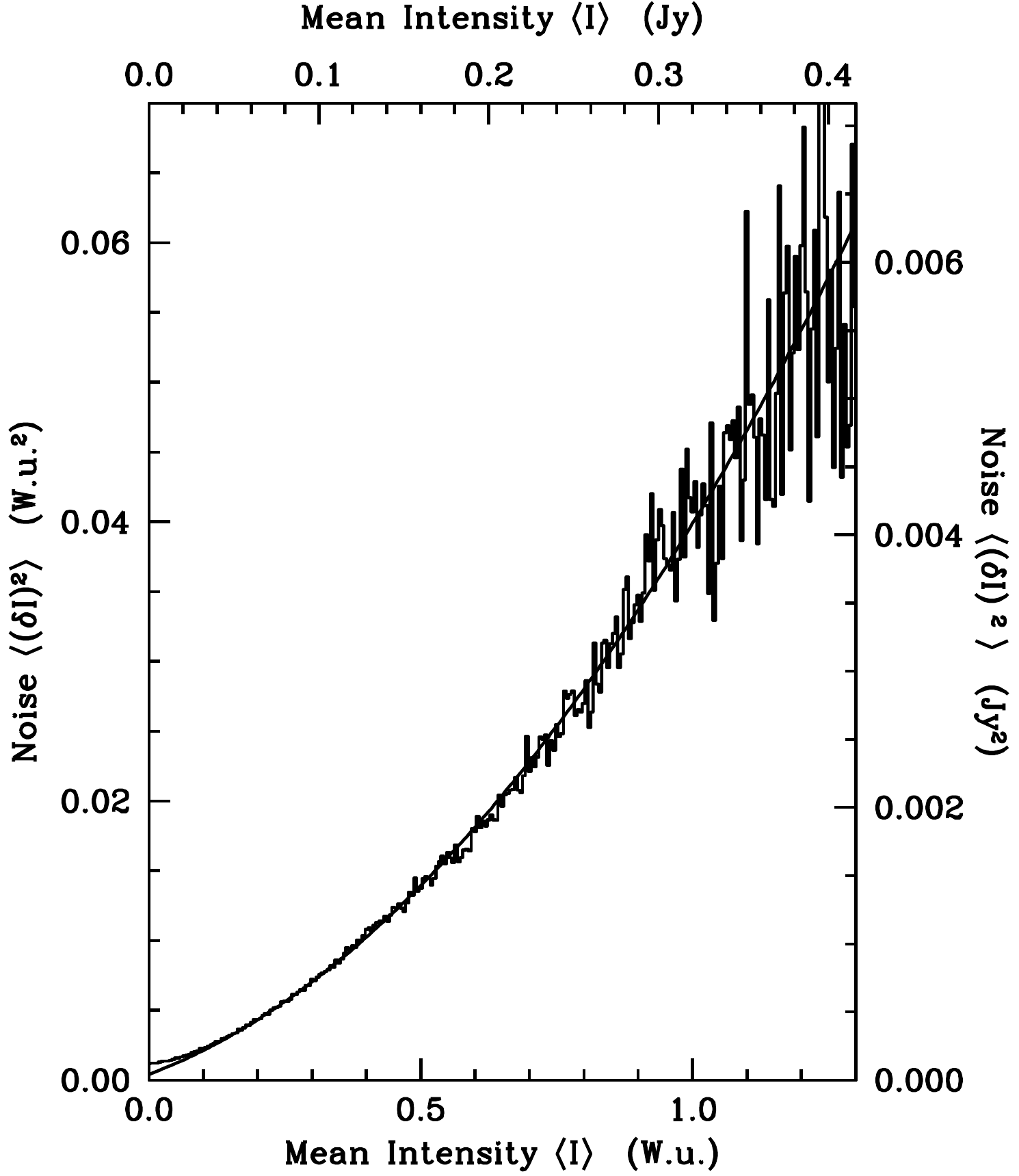}
\figcaption[]{Variation of noise with signal amplitude.
Noise is estimated by differencing consecutive time samples,
amplitude by averaging them.  Smooth curve shows best-fitting parabola of the form of Eq.\ \ref{eq:intensity_noise_polynomial},
for $0.1 < I < 1.2$\ W.u.
\label{fit_sd}}
\end{figure}

We estimate noise by differencing consecutive samples for the single-dish data, as well.
Figure\ \ref{fit_sd} shows the
results.
The effects of self-noise are clearly visible: noise increases at
large signal amplitude, with quadratic dependence, as expected.
For differencing consecutive samples in time, the best-fitting noise
polynomial is, in W.u.: $b_0=4.2\times 10^{-4}$\ W.u.$^2$,
$b_1 = 1.46\times 10^{-2}$\ W.u., and
$b_2=2.5\times 10^{-2}$. 
We obtained this polynomial by fitting the data in Figure\ \ref{fit_sd}
over the range $0.1< I < 1.2$\ W.u..
In Jy, the polynomial coefficients are:
$b_0 = 4.3\times 10^{-5}\ {\rm Jy}^2$,
$b_1= 4.7\times 10^{-3}\ {\rm Jy}$,
and $b_2 = 2.5\times 10^{-2}$.

Because effects of background noise are
relatively small,
we expect that differencing does not fit the values of $b_0$ as well as $b_1$ and $b_2$.
From the sensitivity of Arecibo, we expect
$b_0 = 0.01^2 \cdot (12\ {\rm Jy})^2/(3052\cdot 10\cdot 0.01) = 4.7\times 10^{-5}\ {\rm Jy}^2$,
in perhaps surprisingly good agreement with the fits.
Note that the comparison for $b_0$, unlike $b_2$, involves the calibration
of W.u. to Jy.

Self-noise is particularly important for the WAPP data,
where signal-to-noise ratio is high.
With the duty cycle of the pulsar gate of the WAPP $d_W=1\%$,
using the expression for a gated source of constant flux density in \S\ref{sec:intermediate_term_amplitude_variations},
we find the theoretical value $b_2 =1/(0.01\cdot 3052\cdot 10) = 3.3\times 10^{-3}$.
This should be compared with the fitted value after correction for longer-term amplitude variations,
or $b_2-(\delta A/A)^2 = 1.04\times 10^{-2}$, which is about 3.2 times greater.
Again, note that $b_2$ is dimensionless, so this is independent of calibration.

Values of $b_2$ for both VLBA and WAPP are about 3 times theoretical estimates.
We believe this discrepancy reflects amplitude variations during the 10-sec integration time for one spectrum,
as discussed in \S\ref{sec:intermediate_term_amplitude_variations} above.
The discrepancy for the WAPP is slightly greater.
Our Puschino observations indicate that 
the logarithmic range of amplitudes for a string of pulses, compared at a single phase, is greater near the peak of the pulse, where the Arecibo gate was set,
than in other regions. This could account for the difference.

\subsubsection{Frequency Differences}

In principle, it is possible to estimate noise from adjacent frequency channels in spectra,
if the spectrum is constant over small frequency differences.
However, adjacent channels have deterministic differences in our observations.
These differences are seen 
as structures extending to the largest lags $\tau$ in the secondary spectra,
as discussed in \S\ref{secondary} below.
These structures correspond to the shortest-wavelength structures detectable in our spectra:
those with a period of two spectral channels.
Differences between adjacent channels will measure the intensity in these structures,
as well as noise.
Note that the secondary spectra do not show any structures that extend to the largest rate $\omega$;
thus, we expect that differences between consecutive samples in time will reflect primarily noise.

\subsection{Global Distributions and Fits}\label{sec:global_dist}

\subsubsection{VLBA: Distributions of $Re[V]$ and $Im[V]$}\label{sec:global_dist_vlbi}

\begin{figure}[t]
\epsscale{0.80}
\plotone{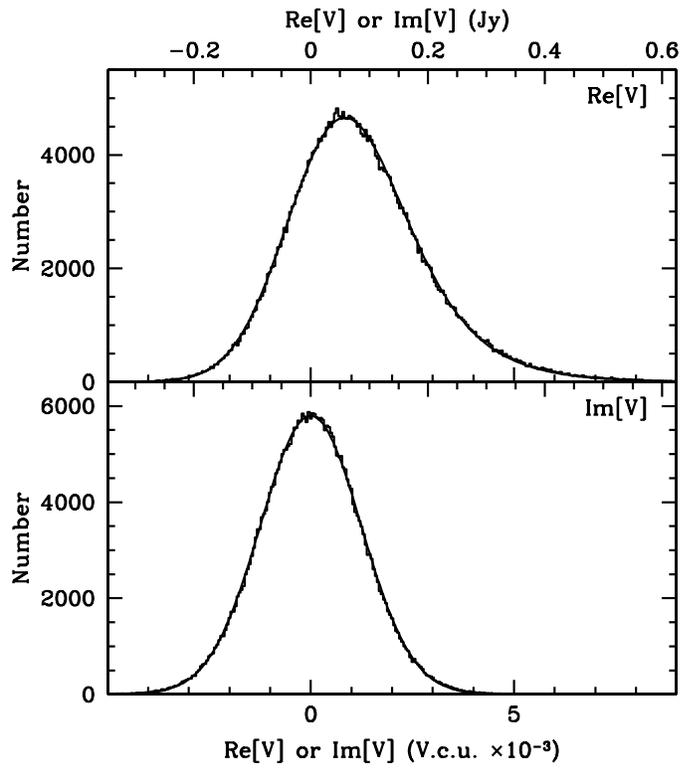}
\figcaption[]{Distributions of real part of interferometric visibility (upper panel) and imaginary part (lower). 
Curves show best-fitting models.  
\label{histo_vlbi}}
\end{figure}

Figure\ \ref{histo_vlbi} shows projections of the 
observed distribution 
of visibility $V(\nu,t)$
onto the real and imaginary axes, and best-fitting models.
The distribution of signal is well-described as an exponential along the positive real axis,
and zero along the negative real axis and elsewhere in the complex plane.
Such an exponential distribution is 
expected for a scintillating source, observed on a short
or zero-length baseline \citep{Gwi01}.
Noise broadens this distribution; each point in the exponential is spread over a Gaussian distribution, centered at that point.
This is not a true convolution because the 
parameters of that Gaussian change as a function of $V$, as Eq.\ \ref{eq:visibility_noise_polynomial} indicates.
We preform the required integral numerically.
Amplitude variations make the observed distribution a sum of scaled copies of exponential distributions,
combined with noise \citep[][\S 2.4]{Gwi00};
we found 
that such distributions, calculated numerically, are indistinguishable from those calculated by simply leaving amplitude variations in the noise coefficient $b_2$,
so we use the simpler results.
The distribution of imaginary part also has an approximately Gaussian distribution,
as the lower panel of Figure\ \ref{histo_vlbi} shows;
here only the linear terms of the noise polynomial contribute.

We use the parameters estimated from differences in \S\ref{sec:diffnoise_vlbi}
for the noise polynomial.
We set the normalization for the model to the number of points observed.
We fitted the scale of the exponential distribution, $I_V$, to the data.
The best-fitting scale is 
$I_V = 1.16\times 10^{-3}$\ V.c.u.
Using the calibration to Jy summarized in \S\ref{sec:calibration_to_Jy} above, 
we find that
this represents an average flux density of $\langle I_V\rangle = 73$\ mJy.
The good fit of the model shows that the bulk of the distribution of visibility is 
consistent with an exponential distribution along the positive real axis, with noise.

\subsubsection{WAPP: Distribution of $I$}\label{sec:global_dist_sd}

\begin{figure}[t]
\epsscale{0.80}
\plotone{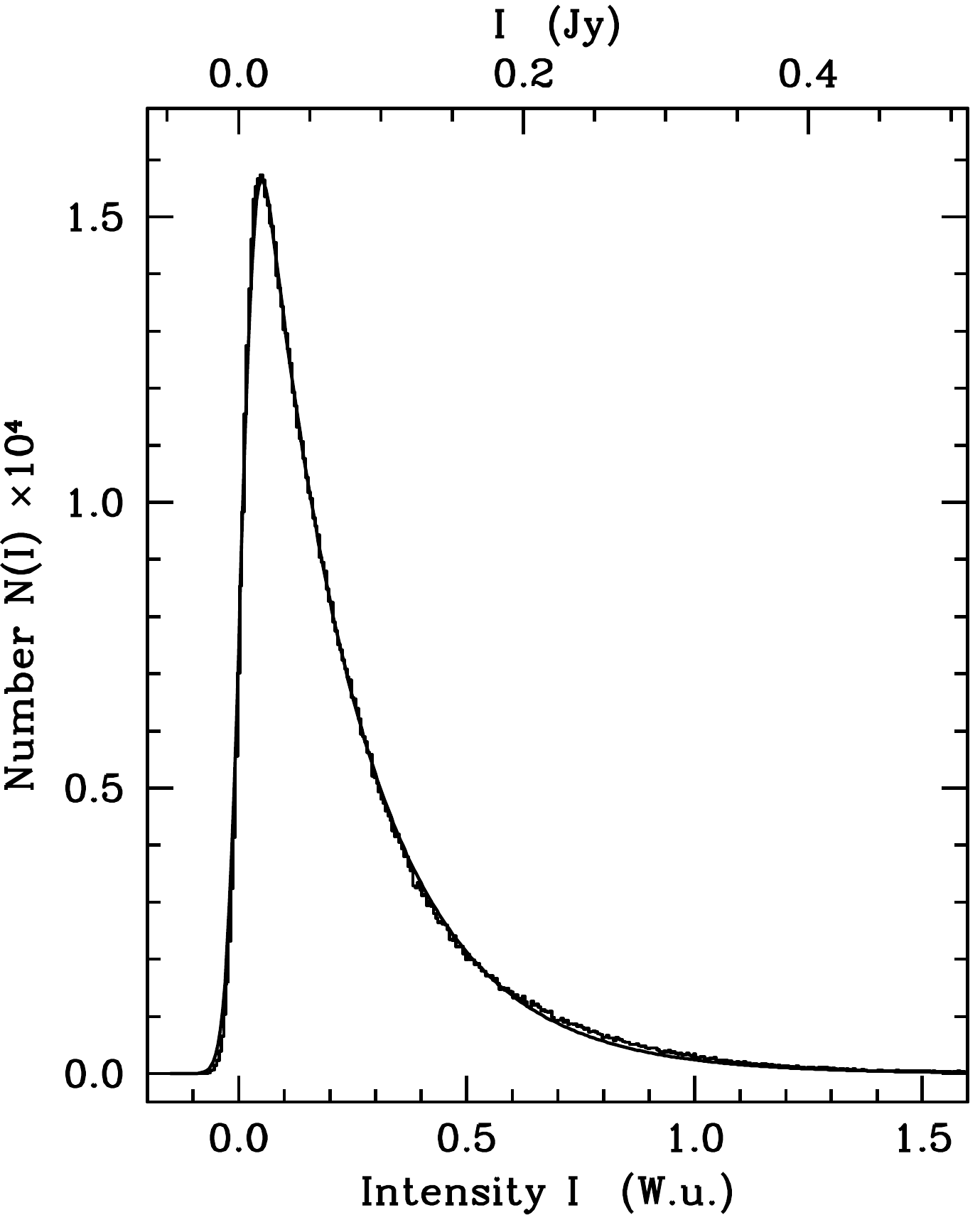}
\figcaption[]{Distribution of intensity of single-dish spectrum.
Smooth curve shows the best-fitting model,
of an exponential with noise and self-noise and an offset (\S\ref{sec:global_dist_sd}).
\label{histo_sd}}
\end{figure}

Figure\ \ref{histo_sd} shows the distribution of intensity
in the dynamic spectrum
for our single-dish WAPP data.  
The distribution is approximately the exponential distribution for $I>0$, expected for the intensity
of a scintillating point source.
A small amount of noise is apparent as a softening of the sharp peak at $I= 0$;
self-noise also affects the distribution at larger intensity.
The distribution departs from the expected smooth tail at large $I$;
because our original dynamic spectrum contains relatively few scintillation elements,
such departures are expected, particularly at large $I$ where occurrences are few.
Noise will tend to smooth the distribution. 
We allowed $b_2$ to parametrize amplitude variations between pulses;
as for the VLBA distributions, a more rigorous treatment led to indistinguishable results.
The best-fitting distribution,
using the model for noise discussed in the previous section,
has exponential scale of $I_W=0.215$\ W.u., or, using the calibration from \S\ref{sec:calibration_to_Jy},
69\ mJy. 
The smooth curve shows this model.
The model also
includes an offset of the exponential from zero intensity; this represents the auto-correlation of system noise,
and any other instrumental contributions to the spectral baseline. 
The fit yields an offset of $I_n = 0.0146$\ W.u.; this is the location of the maximum of the noise-free exponential distribution,
which declines at larger values of $I$ and is zero for smaller values.
The observed distribution extends to even lesser values as a consequence of noise.

Our fitted exponential scale of 69\ mJy
agrees reasonably with the 73\ mJy for the VLBA data above, serving as a check on inter-calibration.
\citet{lor95} found a flux density of 89\ mJy for this pulsar at 408\ MHz;
pulsar flux densities tend to vary with epoch, and their typically-steep spectra can begin to flatten near 300\ MHz,
so we regard this as satisfactory agreement.

\subsection{Noise Inventory}\label{sec:noise_power}

In this section, we inventory the mean square noise in the VLBA and WAPP dynamic spectra
using our fitted parameters,
for comparison with noise in the secondary spectrum via Parseval's theorem.
Although we cannot know the contribution of signal and noise to a particular measurement,
the fits of \S\ref{sec:global_dist} yield the global distribution of signal, and Eqs.\ \ref{eq:visibility_noise_polynomial} and \ref{eq:intensity_noise_polynomial} give the distribution of noise for each signal value.
We can use these to infer the contributions of signal and noise to the mean square intensity in the dynamic spectrum.
The resulting sums involve averages over realizations of both noise and of scintillation,
and are consequently approximate.
The departures of the model fits from the actual distributions in Figures\ \ref{histo_vlbi} and \ref{histo_sd}
give an idea of the error of the estimates.

\subsubsection{Noise for the VLBA Dynamic Spectrum}\label{sec:noise_inventory_VLBA}

For the VLBA data, the estimated mean square noise is 
\begin{eqnarray}
|\delta V|^2 &=\displaystyle{\int}_0^{\infty} dA\, P(A) & \int_0^{\infty} dV\, \left( \textstyle{ {{1}\over{A\, I_V}} \exp\left\{-{{V}\over{A\, I_V}}\right\} } \right)\, \left[ 2 b_0 + 2 b_1 V + b_2 V^2 \right]
\label{eq:vlba_noise_integral}
\end{eqnarray}
In this integral, 
$A$ is the gain-like factor describing amplitude variations over longer times, as defined in \S\ref{sec:long_term_amp_vars_theory};
$P(A)$ gives the distribution of $A$, as presented in Figure\ \ref{fig:sumt}.
The following term,
in parentheses, gives the 
exponential distribution of visibility $V$, in the absence of noise.
The last term, the polynomial in square brackets, gives the noise power at a given $V$.
Factors of 2 represent noise in real and imaginary parts.
We prefer to separate the amplitude variations from $b_2$ for comparison with the secondary spectrum,
but Table\ \ref{noise_table} presents both this result and that of including amplitude variations as a contribution to $b_2$.
We perform the integration numerically, using parameters from fits in \S\ref{sec:diffnoise_vlbi} and\ \S\ref{sec:global_dist_vlbi} above.
Background noise comprises most of the noise budget for the VLBA, so $b_0$ dominates the integral.
Table \ref{noise_table} gives the estimated mean square noise,
calculated using parameters in
Table \ref{parameters_table}.

\subsubsection{Noise for the WAPP Dynamic Spectrum}\label{sec:noise_inventory_WAPP}

An expression similar to Eq.\ \ref{eq:vlba_noise_integral} gives noise for the WAPP.
but using $I$ rather than $V$, and the distribution offset by $I_n = 0.0146\ {\rm W.u.}$:
\begin{eqnarray}
|\delta I|^2 &=\displaystyle{\int}_0^{\infty} dA\, P(A) & \int_{I_n}^{\infty} dI\, \left( \textstyle{ {{\exp\{ I_n/I_W\}}\over{A\, I_W }} \exp\left\{-{{I}\over{A\, I_W}}\right\} } \right)\, \left[ b_0 + b_1 I + b_2 I^2 \right]
\label{eq:wapp_noise_integral}
\end{eqnarray}
Note that the factors of 2 are missing in this case, since the intensity must be real.
Table\ \ref{noise_table} shows the results, and that self-noise dominates the noise budget for the WAPP.  

\subsection{Secondary Spectra: Doppler Rate - Delay Domain}\label{secondary}

\subsubsection{Formation of Secondary Spectra}\label{form_secondary}

\begin{figure}[t]
\epsscale{1.00}
%
\plotone{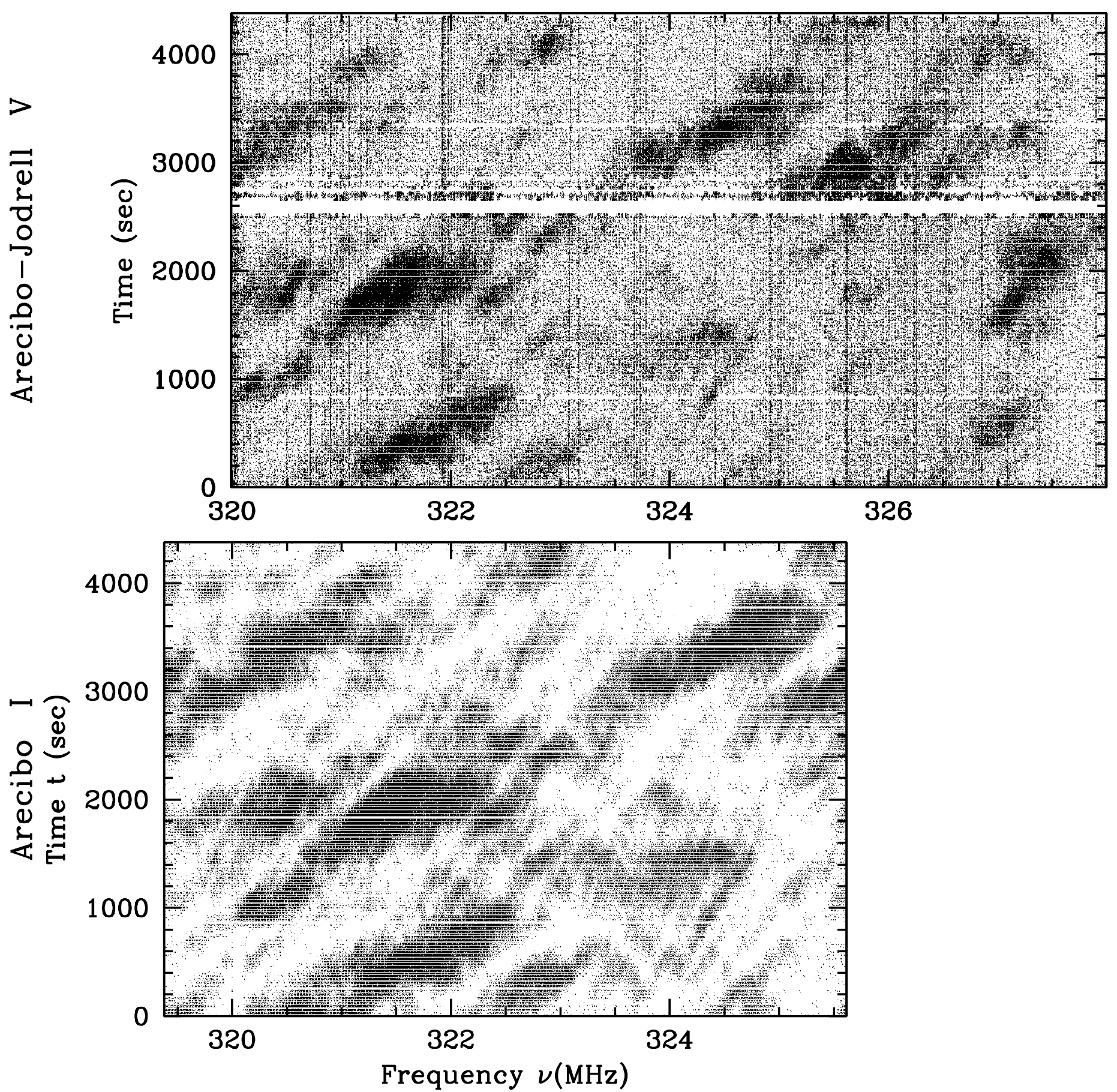}
\figcaption[]{Amplitudes of dynamic spectra for Arecibo (lower) and Arecibo-Jodrell baseline (upper).
The plots span the same time range (vertical dimension) but different ranges
of frequency, as indicated by horizontal offset.
Horizontal gaps in the Arecibo-Jodrell data indicate scan start and end,
or shorter data gaps;
vertical stripes show interference.
Grayscales were adjusted to show similarity of the plots.  
\label{fig:align}}
\end{figure}

We Fourier transformed the VLBA and WAPP data to form lag-rate correlation functions.
The Fourier transform concentrates the power in some features; in particular, it shows the parabolic arcs.
Comparison of the dynamic spectra shows that they are identical, at our noise levels.
The Fourier transform also allows us to estimate noise far from ordered features, and thus form an estimate of uncorrelated
noise.
So as to produce the cleanest spectrum possible for comparisons, we transformed
the first, nearly continuous time segment of data,
for the section of spectrum that overlaps. This comprises the time interval 160 through 2600\ sec
in Figure \ref{fig:align}.
To make the comparison most direct,
we reduced the frequency range of both data sets to the overlapping range in Figure\ \ref{fig:align},
and convolved the narrower frequency channels of the WAPP data to the resolution of the broader VLBA channels.
Before the Fourier transform, 
we set to zero the short time gap in the VLBA data at 830-860 sec visible in Fig\ \ref{fig:align},
and the 72 channels  showing the strongest interference.
In the WAPP data, we set to zero 3 channels showing signs of interference, and one time sample.

\subsubsection{Comparison of Secondary Spectra}\label{sec:compare_secondary}

\begin{figure}[t]
\epsscale{0.9}
\plotone{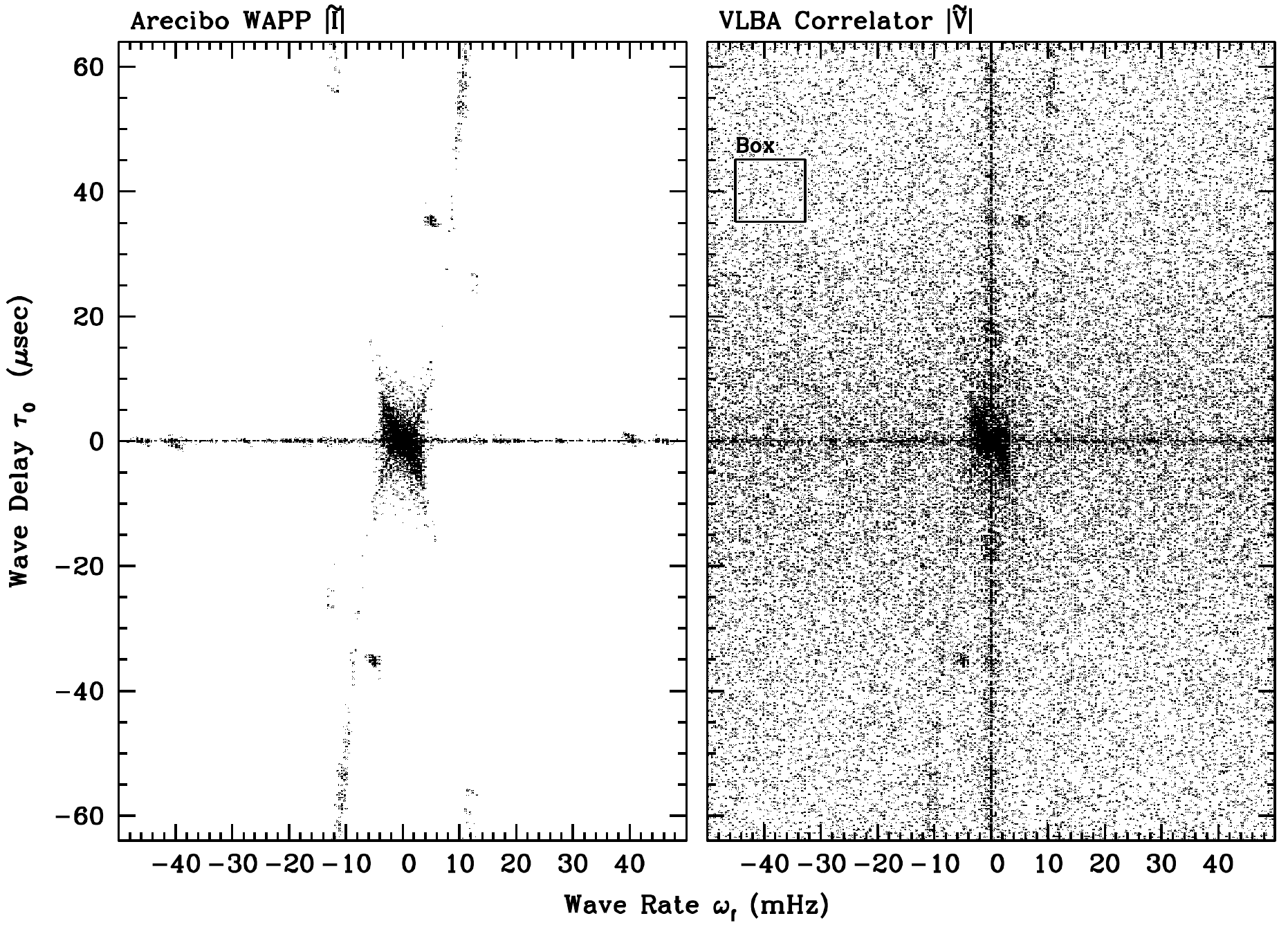}
\figcaption[]{
Secondary spectra for Arecibo single-dish data (left panel)
and Arecibo-Jodrell baseline (right).
Intensity scales are the same in ${\rm Jy}^2$.
Box shows region for noise estimate.
\label{fig:secspecs}}
\end{figure}

Figure\ \ref{fig:secspecs} compares secondary spectra for the two instruments.
It displays the amplitude of the
VLBA and WAPP data,
on the same Jy scale. 
For single-dish data such as the WAPP data,
values at $(\tau_0,\omega_0)$ and 
$(-\tau_0,-\omega_0)$ must have identical amplitude (and opposite phase);
for the VLBA data this need not be the case; so we display both.
However, as the plot shows, the conjugate range appears qualitatively the same for the VLBA data.
(Note that $\tau$ is given first, but is displayed along the vertical axis;
here we follow the convention of previous work).
The regions at larger $\pm\omega$ are consistent with nearly stationary noise, as we discuss below.
Clearly, the background noise is much higher for the VLBA data.
Indeed, the background noise level for the secondary spectrum of the WAPP is invisible on the scale of Figure\ \ref{fig:secspecs}.

\begin{figure}[t]
\epsscale{0.9}
\plotone{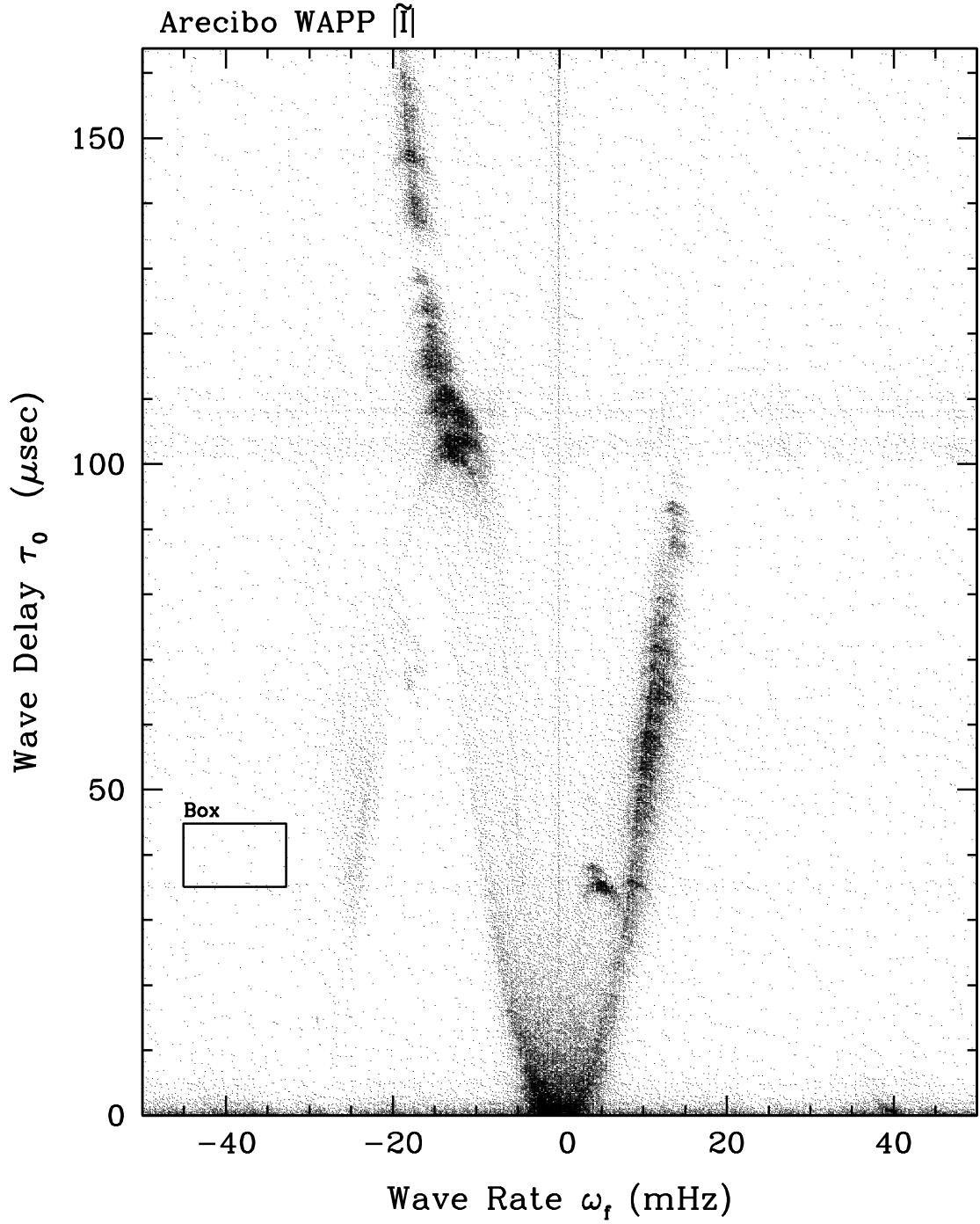}
\figcaption[]{
Secondary spectra for Arecibo single-dish data.
Box shows region for noise estimate.
\label{fig:wappsec}}
\end{figure}

We show the WAPP spectrum to a more sensitive level in Fig.\ \ref{fig:wappsec}.
The WAPP data are uninterrupted by time gaps or interference, so we form the secondary spectrum from the full data set, yielding a larger range of $\tau$
and finer resolution in $\omega$ than Fig.\ \ref{fig:secspecs}, which includes only the region of overlap.

A source of constant flux density would produce a spike at the origin of the secondary spectrum.
Scintillation broadens this spike, over a region corresponding to the time- and frequency scales of scintillation,
as estimates in \S\ref{scint_params} above.
A parabolic arc, with somewhat asymmetric arms, is visible in 
both secondary spectra, although it is clearer in the WAPP data because the noise is less.
The arc is well approximated by a parabola with 
curvature $0.50\ {\rm \mu sec}/{\rm mHz}^{2}$.
for both VLBA and WAPP of observations.  
The parabolic arc extends to the edge of the secondary spectrum,
limited by the spectral resolution of our data.  
\citet{Bri10} observe this arc with much higher spectral resolution, and obtain a consistent value for the curvature.
An isolated ``clump'' of emission appears in WAPP and VLBA spectra,
near $(\tau_0,\omega_0) = (35\ \mu{\rm s}, 5.0\ {\rm mHz})$,
and likewise at $(-\tau_0,-\omega_0)$.
Brisken et al. do not observe this clump; 
features of this type tend to last only a short time \citep{Hil03}.

The arc and clump have small spectral power, relative to the power in larger-scale scintillation.
The spectral power in the larger-scale scintillation is 
$1.12\times 10^{-1}\ {\rm W.u.}^2$,
within
an ellipse centered at the origin of the secondary spectrum, with semimajor axes of $2\Delta\nu_d$ and $2t_d$.
The power in the 
arc is about $2.7\times 10^{-3}\ {\rm W.u.}^2$, outside of this ellipse.
About half of the power in the arc is close to the origin, at $\tau<40\ \mu{\rm sec}$.
The spectral power in the clump is about $1.7\times 10^{-4}\ {\rm W.u.}^2$.
The background noise as estimated from the box in \S\ref{sec:noise_box} below does not contribute significantly to these measures.
We can not measure the spectral power outside our maximum delay; some of this may be aliased into the observed region.
In any case, the spectral power in the arc is about 3\% of the total, and that in the clump is an order of magnitude less.

Lines appear along the horizontal rate axis of both spectra, and along the vertical delay axis of the VLBA spectrum.
The vertical line corresponds to effects that are constant with time, but vary with frequency.
These most likely represent effects of editing to remove the interference spikes visible in 
the dynamic VLBA spectrum, Figure\ \ref{fig:align}.
The line along the rate axis represents effects that change with time but are constant with frequency:
amplitude variations of the pulsar probably contribute most to these.
In the dynamic spectrum, amplitude variations multiply signal by $A(t)$;
thus, in the correlation function in the lag-rate domain, signal is convolved with the Fourier transform of $A(t)$.
This transform is a broad function of rate $\omega$, but a delta-function in delay $\tau$.
Its effects are visible in Figure\ \ref{fig:wappsec} as a horizontal broadening of the primary scintillation maximum, as well as of some of the outlying structures associated with the arc.

\subsubsection{Noise from Box}\label{sec:noise_box}

To estimate noise, we measured the mean square flux density in a box far from recognizable structures.
Fig.\ \ref{fig:wappsec} shows this ``noise box''.
The box extends over 
 $35<\tau<45\ \mu{\rm sec}$ and $-45<\omega<-32\ {\rm mHz}$.
For the VLBA observations, we used the secondary spectrum formed from the region of overlap of the two data sets, and 
uninterrupted time sampling.
The root-mean-squared noise in the box is $4.0\times 10^{-6}\ {\rm V.c.u.}^2$.
The summed, squared noise extrapolated from the box over the entire spectrum, $N_2 = N_s\times N_t = 720\times 244$ samples,
is $2.8\times 10^{-6}\ {\rm V.c.u.}^2$.
This agrees reasonably well with the mean square noise in \S\ref{sec:noise_inventory_VLBA} above,
as Table\ \ref{noise_table} summarizes.
This strengthens our confidence in the noise model and the procedure for the comparison.

We also found the noise for the WAPP from the box shown in Fig.\ \ref{fig:wappsec}. 
The mean squared noise in the box is $1.88\times 10^{-9}\ {\rm W.u.}^2$.
Extrapolated to the entire secondary spectrum, of $N_2 = 2048\times 433$ samples,
this corresponds to summed squared noise of $1.69\times 10^{-3}\ {\rm W.u.}^2$.
As in Table\ \ref{noise_table} shows, this noise level is a factor of 3.7 smaller than that inferred 
using time-differencing, and a factor of 2.8 smaller when effects of amplitude variations are removed.
The noise level is a factor of 4 greater than expected from background noise alone, as parametrized by $b_0$.
Whereas noise for the VLBA appears as a uniformly-distributed background level in the secondary spectrum, noise for the WAPP appears to be
distributed unevenly, and to be concentrated along with signal.

Self-noise dominates the noise budget of the WAPP.
It appears in estimates made by differencing in time, and in the distribution of intensity.
However, it does not appear at the expected level, as a uniformly-distributed background, in the secondary spectrum.
This suggests that only a fraction of self-noise appears as a stationary background in the secondary spectrum.
Correlation of the noise among samples in the dynamic spectrum can redistribute noise 
in the secondary spectrum;
such correlations can arise from intermittent emission,
as discussed in \S\ref{sec:short-term_amplitude_variations} above and in more detail in \citet{Gwi10}.

\section{SUMMARY AND DISCUSSION}\label{sec:summary}

We observed PSR\ 0834$+$06 with two instruments: the VLBA correlator, with data from the Arecibo-Jodrell Bank baseline;
and the WAPP, a single-dish spectrometer at Arecibo.
The observed dynamic spectra agree well, although noise levels are different.
We compared noise levels as estimated in different ways,
and with the different instruments, as summarized below.
In the frequency-time domain of the dynamic spectrum,
we find that noise increases quadratically with signal, as expected because of the contribution of self-noise.
We find that self-noise is greater than expected. We suggest that this discrepancy arises from variations
of the intrinsic flux density of the source on intermediate timescales, of $300\ \mu{\rm sec}$ to 10\ sec.
We also find that noise is not uniformly distributed with lag and rate in the secondary spectrum;
we suggest that this indicates correlation of self-noise, because of variations in intrinsic flux density of the source 
on short timescales, $<300\ \mu{\rm sec}$.

\subsection{Noise in the Dynamic Spectrum}

In the time-frequency domain of the dynamic spectrum,
we estimated noise by differencing consecutive spectra.
We estimate noise by differencing observations at consecutive times.
Noise increases with flux density of the source, as expected from self-noise or source noise.
The functional form of the change in noise with the intensity of the source is in good agreement with the theoretical form:
a quadratic polynomial, with the interferometric observations showing only the linear terms in that polynomials out of phase with the signal.
The constant terms $b_0$ in the noise polynomials for the two instruments, describing background noise, are in good agreement with estimates based on antenna sensitivity, 
after calibration.  
The quadratic coefficient $b_2$, describing self-noise, is larger than the expected amount by a factor of about 3, for both VLBA and WAPP,
after correcting for variations of the flux density of
the source between spectra.
We suggest that this excess arises from variation of the flux density of the pulsar on intermediate timescales, 
during the 10-sec period of integration of the spectra.
Such variations are to be expected from the documented variability of this pulsar
\citep{Tay75,Rit76,ad05,ran07}.  

We find that the Dicke Equation must be modified if the source varies during integration of the spectrum.
Such variations lead to an increase in noise.
For a source that is either ``on'' at a particular intensity or ``off'', 
reduction of the number of samples by the duty cycle gives the correct expression for self-noise (\S\ref{sec:intermediate_term_amplitude_variations});
for sources with more complicated amplitude variation, or to take both self-noise and background noise into account,
more complicated expressions are needed \citep[see][\S 3.3.3]{Gwi10}.
Our results are consistent with a duty cycle for the pulsar of about 1/3, on top of the average duty cycle of the pulsar of 1.8\%.
Our formalism can be easily extended to cover variations more complicated than simple variations in time:
for example, variations in both frequency and time.  
In this work, amplitude variations in time appear to be responsible for the increase in noise, since
we determine noise by differencing in time (\S\ref{sec:diffnoise}).

\subsection{Noise in the Secondary Spectrum}

We observe primary scintillation, and the scintillation arc as reported by others \citep{Hil03,Bri10}.
We find that about 97\% of spectral power resides in the elliptical region defined by the pulsar's typical scintillation bandwidth and timescale;
an extended scintillation arc contains about 3\%.
We find that the background noise level in the lag-rate domain represents 95\% of the noise in the dynamic spectrum for the VLBA,
as expected for uncorrelated noise in the dynamic spectrum.
The background noise level in the lag-rate domain represents only 35\% of the noise in the dynamic spectrum for the WAPP.
This indicates that the noise for the WAPP is concentrated near signal in the lag-rate domain, and correlated among samples in the dynamic spectrum.
Because WAPP noise is dominated by self-noise, 
we conclude that samples of self-noise are correlated in the dynamic spectrum.
Such correlation can be caused by intermittent emission by the pulsar, on timescales shorter than the time for accumulation of one spectrum,
$\approx 300\ \mu{\rm sec}$, as we discuss elsewhere \citep{Gwi10}.
Such short-term variations 
have been reported by \citet{Kar78}, who observed that periodic structures with timescales of 160 to $700\ \mu{\rm sec}$
are common in pulsar B0834$+$06.
Such variations are common for other pulsars as well \citep{Han71,Cog96,Jen01,Pop09}.

\subsection{Applications and Further Work}

Our conclusions are important in situations where the contribution of self-noise is significant,
and when the source varies within the time for accumulating one realization of the spectrum.
Typically, this requires that the maximum strength of the source be comparable to the system-equivalent flux density of the telescope.
Here, the maximum strength of the source is the average flux density divided by the actual duty cycle, or more.
Among radio sources, only pulsars and the Sun are known to show variability on these short timescales.

Noise sets important limits for many types of observations.
In this paper, we focus on the spectrum, particularly the dynamic spectrum;
however, noise sets limits in astrometry and pulsar timing.
In astrometry, positional accuracy is approximately instrumental resolution, divided by signal-to-noise ratio:
thus, we expect that variability, and consequent increase in self-noise, may limit astrometric accuracy for strong pulsars.
Similarly, in pulsar timing, accuracy is approximately the width of the narrowest feature in the pulse, divided by signal-to-noise ratio.
Variability within a stable envelope will increase self-noise and degrade timing accuracy,
although narrow features with predictable arrival times could potentially offer greater accuracy.

We observe that the background noise level in the lag-rate correlation function decreases when the source varies within the time for accumulating one 
realization of the spectrum.
Interestingly, this reduction is independent of the form of the propagation kernel,
as long as the accumulation time exceeds the duration of the kernel \citep{Gwi10}.
The noise measures the fourth moment of electric field, so it is less sensitive than 
techniques that use the intensity, the second moment.
Nevertheless, it offers the possibility of detecting or measuring intrinsic variability, in cases where the propagation kernel is unknown,
or is impossible or impractical to invert.

New instrumentation, now available, offers the possibility of testing and extending our conclusions.
The DiFX software correlator now operational at the VLBA and other interferometers offers the 
possibility of high-resolution spectra for both cross-power and autocorrelations, along with flexible pulsar gates \cite{Del07}.
Baseband observing systems offer similar capabilities, and the possibility of examining statistics of individual realizations of the spectrum in parallel with
studies of source variability \citep{van10}.
Such studies can further extend understanding of the role and uses of noise in radioastronomical observations.

\acknowledgments

We wish to thank T. Ghosh and C. Salter for assistance with amplitude calibration,
M.A. Walker for stimulating discussions, R. Ramachandran for help with observations,
and the VLBA analysts for patiently correlating data in an awkward mode.
We thank an anonymous referee for helpful suggestions.
C.R.G. and M.D.J. thank the National Science Foundation (AST-1008865) for financial support.
T.V.S. thanks the Russian Foundation for Basic Research (project code 0902-00530) for financial support.

\clearpage

\newpage
\clearpage

\newpage
\clearpage

\newpage
\clearpage

\newpage
\clearpage

\newpage
\clearpage

\newpage
\clearpage

\newpage 
\clearpage

\newpage
\clearpage

\newpage
\clearpage

\newpage
\clearpage

\begin{deluxetable}{lllll}
\tablenum{1}
\tablecolumns{4}
\tablewidth{0pc}
\tablecaption{Parameters of Distributions}
\tablehead{
\colhead{Quantity}&\colhead{Symbol}&\colhead{VLBA}&\colhead{WAPP} 
}
\startdata%
\sidehead{Noise Polynomial}
Time Differencing              & $b_0$      & $1.52\times 10^{-6}\ {\rm V.c.u.}^2$  & $4.2\times 10^{-4}\ {\rm W.u.}^2$  \\
                                          &$b_1$       & $1.88\times 10^{-5}\ {\rm V.c.u.}$      & $1.46\times 10^{-2}\ {\rm W.u.}$     \\
                                          &$b_2$       &   0.0163                                               & 0.025                                                \\
Longer-term Amplitude Variations         &$(\delta A/A)^2$ & 0.0144                                                   &0.0144                              \\
\sidehead{Other Parameters}
Signal: Exponential           &$I_{V},I_{W}$    & $1.16\times 10^{-3}\ {\rm V.c.u.}$       & 0.215\ {\rm W.u.}                               \\ 
Offset                                &$I_n$                    & -                                                          & 0.0146\ {\rm W.u.}     \\
Calibration                         &                        &  63\ Jy/V.c.u.                                       & 0.32\ Jy/W.u.              \\
\enddata
\label{parameters_table}
\end{deluxetable}

\clearpage
\begin{deluxetable}{lcc}
\tablenum{2}
\tablecolumns{3}
\tablewidth{0pc}
\tablecaption{Noise Comparisons}
\tablehead{
\colhead{}&\colhead{VLBA}&\colhead{WAPP} \\
\colhead{}&\colhead{$({\rm V.c.u.})^2\times 10^{-6}$}&\colhead{$({\rm W.u.})^2 \times 10^{-3}$}
}
\startdata
\sidehead{Mean Square Noise in Dynamic Spectrum}
Noise from Fitted Parameters                                                & $3.1$ &  $6.2\phantom{3}$\\
Noise from Fitted Parameters Corrected Amplitude Variation       & $3.0$ &  $4.8\phantom{3}$\\
Background Noise $b_0$ Only                                & $3.0$ &  $0.42$\\
\sidehead{Total Uniformly-Distributed Noise in Secondary Spectrum}
Mean Square Noise in Box $\times {N_{2}}^{a}$              & $2.8$ & $1.69$ \\ 
\enddata
\tablenotetext{a} {Elements in Secondary Spectrum: $N_{2}=({\rm spectral\ channels})\times ({\rm time\ periods})$. }
\label{noise_table}
\end{deluxetable}

\end{document}